\documentclass[12pt]{article}
\input xy
\xyoption{all}
\input epsf.tex

\usepackage{graphicx}
\usepackage{mathrsfs}
\usepackage{cancel}
\usepackage{amssymb,amsmath}
\usepackage{epstopdf}
\usepackage[top=2.5cm,bottom=3cm,left=2.5cm, right=2.5cm]{geometry}

\begin{document}

\begin{center}
\Large{\bf Entangled spinning particles in charged and rotating black holes}\\
\vspace{1cm}

\large Felipe Robledo-Padilla$^a$\footnote{electronic address: {\tt
felipe.robledopd@uanl.edu.mx}}, Hugo
Garc\'{\i}a-Compe\'an$^b$\footnote{electronic address: {\tt
compean@fis.cinvestav.mx}}
\\
[2mm]
{\small \em $^a$Facultad de Ciencias F\'{\i}sico-Matem\'{a}ticas}  \\
{\small \em Universidad Aut\'{o}noma de Nuevo Le\'{o}n} \\
{\small \em Ciudad Universitaria, San Nicol\'{a}s de los Garza} \\
{\small \em Nuevo Le\'{o}n, 66450, M\'{e}xico} \\
[2mm]
{\small \em $^b$Departamento de F\'{\i}sica} \\
{\small \em  Centro de Investigaci\'on y de Estudios Avanzados del IPN}\\
{\small\em P.O. Box 14-740, 07000 M\'exico D.F., M\'exico}\\

\vspace*{2cm}
\small{\bf Abstract}
\end{center}

\begin{center}
\begin{minipage}[h]{14.0cm} {Spin precession for an EPR pair
of spin-1/2 particles in equatorial
orbits around a Kerr-Newman black hole is studied. Hovering
observers are introduced to ensure fixed reference frames in order
to perform the Wigner rotation. These observers also guarantee a
reliable direction to compare spin states in rotating black holes.
The velocity of the particle due frame-dragging is explicitly
incorporated by addition of velocities with respect the hovering
observers and the corresponding spin precession angle is computed.
The spin-singlet state is observed to be mixed with the spin-triplet
by dynamical and gravity effects, thus it is found that  a perfect
anti-correlation of entangled states for these observers is
deteriorated.  Finally, an analysis concerning the different limit
cases of parameters of spin precession including the frame-dragging
effects is carried out.}
\end{minipage}
\end{center}

\bigskip
\bigskip


\vspace{3cm}

\leftline{\today}
\newpage

\section{Introduction}

Entanglement of quantum states has taken a great deal of attention
as a fundamental issue in physics since Einstein-Podolsky-Rosen
(EPR) challenged quantum mechanics as a complete model to describe
reality \cite{EPR}. With the work by Bohm-Aharanov \cite{B-A} for
spin-entangled particles and Bell's hidden variables \cite{Bell} it
was possible to enforce that quantum mechanics is the correct
description of the quantum realm and eventual experimental results
\cite{C-F, Aspect,Tittel} confirmed this fact. In recent years a
great deal of research on entangled states has been focused on
quantum communication and teleportation
\cite{popescu,bennet,brassard}, quantum computation
\cite{jozsa,r-b,g-c,knill_etal} and quantum cryptography
\cite{ekert,bennet_etal}.

More recently the entanglement behavior of quantum states in
classical gravitational fields has been studied in the literature.
The first steps were taken in the context of special relativity
\cite{a-m,t-u_sr1,r-s,g-a,ahn_etal, l-d} and later there were
integrated within the framework of general relativity for the
Schwarzschild spacetime \cite{T-U} and for the Kerr-Newman spacetime
\cite{S-A}.

In particular for the case of the Schwarzschild spacetime  Terashima
and Ueda \cite{T-U} considered a pair of spinning particles in an
entangled state moving on equatorial motion. Their results showed
that the acceleration and the gravitational effects deteriorate the
EPR correlation in the direction that are the same than in
non-relativistic theory, and apparently decrease the degree of the
violation of Bell's inequality. They also found that near the event
horizon there exists even a small uncertainty in the identification
of the positions of the observers leading to a fatal error in
identifying the measurement direction needed to maintain the perfect
EPR correlation, which is due an extremely rapid spin precession.
This implies that the choices of four-velocity vector and the
vierbein (or tetrad) are important for non-local communication in a
curved spacetime using an EPR pair of spins \cite{T-U}.

The case of a rotating and charged black hole was studied in
\cite{S-A}. There were considered an observer at infinity and a free
falling observer and the it is found that the EPR correlation is unmeasurable for
both cases at the event horizon and below. The spin precession
approaches negatively to infinity and that result implies an impossibility to extract the EPR correlation in that region.

The aim of this paper is to  extend previous work \cite{T-U,S-A} by
considering a different kind of observers (not at infinitye) and by
including the frame-dragging effects explicitly. For this, we consider
Kerr-Newman spacetime in different coordinate systems from that it was
used in \cite{S-A} and then we study the effects in the different limit cases.

The approach we used in the present paper follows Terashima and Ueda
\cite{T-U} in analysis, notations and conventions. The main idea is
to Lorentz transform the spin quantum state, which is locally well
defined in the non-relativistic theory. This transformation must
preserve quantum probabilities of finding the spin state in the
particular direction measured on a local inertial frame. In order to
guarantee this, the transformation changing quantum state from a
point to another one, must be unitary. The Wigner rotation matrix
\cite{Wigner} achieves it. This rotation is composed by
infinitesimal Lorentz transformations, which consist of a boost
along the radial direction and a rotation in the angle direction of
the orbital particle. Finally, it is found a precession of spin of a
particle moving in curved spacetime due the acceleration of the
particle by an external force and due to the difference between
local inertial frames at different points.

This paper is organized as follows. Section \ref{review} is an
overview of the Kerr-Newman spacetime, its particular metric, their
local inertial frames and event horizons. In Section
\ref{frame-dragging} the frame-dragging is discussed and hovering
observers are introduced. Section \ref{precession} formulates step
by step the spin precession in circular orbits on equator and
Section \ref{EPR_correlation} calculates the EPR correlation by
Wigner rotations due the motion and dragging velocity of each
particle over the rotating spacetime. The relativistic addition of
velocities is performed  by the introduction of the zero angular
momentum observers (ZAMOs) as a preliminary step. Our results are
discussed in Section \ref{results} in terms of the limiting values of the different parameters. Then Reissner-Nordstr\"{o}m, Kerr and Kerr-Newman cases
are analyzed independently and with these results the Bell's
inequality is analyzed, closing the section. Conclusions and final
remarks are presented in Section \ref{conclusions}.

\section{Rotating and Charged Black Holes: The Kerr-Newman geometry}

\label{review} The Kerr-Newman solutions form a three-parameter
family of spacetime metrics, which in in Boyer-Lindquist coordinates
\cite{Wald,R-H} is given by
\begin{equation}
    ds^2 = -\frac{\bar{\Delta}}{\Sigma}( dt-a \sin^2\theta d\phi)^2 + \frac{\Sigma}{\bar{\Delta}}dr^2 +\Sigma d\theta^2 + \frac{ \sin^2 \theta}{\Sigma} [adt- (r^2+a^2)d\phi]^2,
    \label{eq:metric}
\end{equation}
where
\begin{equation}
\begin{array}{l}
\bar{\Delta} = r^2 - 2m r + a^2+e^2, \\
\Sigma = r^2 + a^2\cos^2\theta .
\end{array}
\label{eq:deltasigma}
\end{equation}

The three parameters of the family are electric charge $e$, angular
momentum $a$  and mass $m$. The spacetime metric is expressed in
geometric units (\emph{G = 1} and \emph{c = 1}).

When $\bar{\Delta} \to 0$ the metric coefficient $g_{rr} \to
\infty$, then Eq.~(\ref{eq:metric}) becomes problematic and the
metric fails to be strongly asymptotically predictable, and thus it
does not describe black holes \cite{Wald}. Therefore, this metric
has physical meaning when $a^2+e^2 \leq m^2$, which is consequence
of solving $\bar{\Delta} = r^2 - 2m r + a^2+e^2 =0$ in $g_{rr}$.
This component of the metric establishes two possible values of $r$
for the Kerr-Newman black holes
\begin{equation}
r_\pm=m \pm \sqrt{m^2-a^2-e^2},
\label{eq:horizons}
\end{equation}
whose horizon is denoted by $r_+$. As in the Schwarzschild case, $r > r_+$
is the region where we can obtain sensible causal information of the
system.

An important difference in Kerr-Newman is that the horizon is below the Schwarzschild radius $r_\mathrm{s} = 2m$, as can be seen from $r_+$
equation. When $a^2+e^2= m^2$ it is called extreme Kerr-Newman black
hole, hence $r_+=r_-$ and the horizon is placed at $r=m$.

In order to describe the motion of spinning particles in a curved
spacetime, the local inertial frame at each point is defined by a
vierbein chosen as \cite{R-H}:
\begin{equation}
\begin{array}{l}
e_0{}^\mu(x) = \left( \displaystyle \frac{r^2+a^2}{\sqrt{\bar{\Delta} \Sigma}}, 0, 0, \displaystyle \frac{a}{\sqrt{\bar{\Delta} \Sigma}} \right), \\
e_1{}^\mu(x) = \left( 0, \displaystyle \sqrt{\frac{\bar{\Delta}}{\Sigma}}, 0, 0 \right), \\
e_2{}^\mu(x) = \left( 0, 0, \displaystyle \frac{1}{\sqrt{\Sigma}}, 0 \right), \\
e_3{}^\mu(x) = \left( \displaystyle \frac{a\sin \theta}{\sqrt{\Sigma}}, 0, 0, \displaystyle \frac{1}{\sqrt{\Sigma}\sin \theta}\right).
\end{array}
\label{eq:veirbein}
\end{equation}

It is easy to show that this vierbein satisfies the standard conditions \cite{nakahara}:
\begin{equation}
\begin{array}{r}
e_a{}^\mu(x)e_b{}^\nu(x)g_{\mu \nu}(x) = \eta_{ab}, \\
e^a{}_\mu(x)e_a{}^\nu(x) = \delta_\mu{}^\nu, \\
e^a{}_\mu(x)e_b{}^\mu(x) = \delta^a{}_b,
\end{array}
\label{eq:conditions}
\end{equation}
where the Latin indices run over the flat Lorentz indices 0,
1, 2, 3; the Greek indices run over the four general-coordinate
labels \emph{t, r,} $\theta$, $\phi$ and Einstein's sum convention
on the repeated indices is holding.

In Ref. \cite{S-A} a similar analysis was carried out, but a
different vierbein was chosen, where the frame-dragging effects were
not explicitly obtained.

\section{Frame-dragging}

\label{frame-dragging}

Consider a freely falling test particle with four-velocity $u^\mu$ in the exterior of Kerr-Newman black hole. The covariant component of a four-velocity in a direction of a given symmetry is a constant. For an observer at infinite, they are two conserved quantities: the relativistic energy per unit mass $E=-u_\phi$ and the angular momentum per unit mass $L_z=u_\phi$.

Because $g_{\mu\nu}$ is independent of $\phi$, the trajectory of the particle still conserves angular momentum $u_\phi$.  But the presence of $g_{t \phi} \neq 0$ in the metric introduces an important new effect on the particle trajectories \cite{schutz}. The freely fall test particle will adquire angular momentum as it is approaching the black hole. To see that, consider the contravariant four-velocity for a test particle, which  is
\begin{equation}
\begin{array}{l}
  \displaystyle \frac{dt}{d\tau} = u^t = g^{tt}u_t + g^{t\phi}u_\phi, \\
  \displaystyle \frac{d\phi}{d\tau} = u^\phi = g^{t\phi}u_t + g^{\phi \phi}u_\phi.
\end{array}
\label{eq:contravariant_particle_velocity}
\end{equation}

This test particle would be falling now from infinite with zero angular momentum, i.e. $u_\phi=0$. Despite the fact that initially the particle falls radially with no angular momentum, it acquires an angular motion during the infall \cite{raine-thomas}, that is, from
(\ref{eq:contravariant_particle_velocity}) the angular velocity as seen by a distant observer is
\begin{equation}
 \omega(r,\theta)=\frac{d\phi}{dt} = \frac{d\phi/d\tau}{dt/d\tau}
 = \frac{u^\phi}{u^t}=\frac{ g^{t\phi} }{ g^{tt} }.
 \label{eq:omega_formula}
\end{equation}

Last equation means that a rotating relativistic body influences the surrounding matter in
addition directly through its rotation. Thus a particle dropped in a
Kerr-like black hole from infinite is dragged just by the influence of
gravity so that it acquires an angular velocity $\omega$ in the same
direction of rotation of the black hole. This effect weakens with distance
\cite{schutz}. From a physical point of view we can interpret this
phenomenon as a dragging round of the local inertial frames of
reference by the rotating hole. This inertial frame rotates
with an angular velocity $\omega$ relative to infinite, hence is
dragged round with the rotation of the hole \cite{raine-thomas}.

Consider now the same particle in circular orbit around
the rotating black hole ($u_r = u_\theta = 0$). From
(\ref{eq:contravariant_particle_velocity}) we get
\begin{equation}
 u^\phi_{fd} = -\frac{g_{t\phi}}{g_{\phi \phi}}u^t_{fd} = \omega u^t_{fd},
\label{eq:uphi_fd}
\end{equation}
where we now identify $u^\mu_{fd}$ as the velocity of the particle due this frame-dragging. Using the normalization condition for velocities $u^\mu u_\mu =
-1$, it can be shown that
\begin{equation}
u^t_{fd}=\sqrt{ \frac{ -g_{\phi \phi} }{ g_{tt}g_{\phi \phi} - (g_{t\phi})^2 } }.
\label{eq:ut_fd}
\end{equation}

Both Eqs. (\ref{eq:uphi_fd}) and (\ref{eq:ut_fd})  constitute the components of the four-velocity of a test particle due the frame-dragging as seen by a distant observer in the general frame.

We shall see later that the spin precession angle is calculated by infinitesimal Lorentz transformations of the velocity of a particle in a local inertial frame, because the spin is only defined in this kind of frames.

Then, in order to find the velocity of a particle in a local inertial frame, we will adopt a convenient set of observers that ``hovered'' at fixed coordinate position. But first, as seen by long distances observers, the contravariant four-velocity is
\begin{equation}
 u^\mu_h = (dt/d\tau,~0,~0,~0)=((-g_{tt})^{-1/2},~0,~0,~0),
\label{eq:hovering_velocity_formula}
\end{equation}
and their covariant four-velocity is obtained by lowering indices, that is
\begin{equation}
 u_{\mu_h} = \left( -\sqrt{-g_{tt}},~0,~0,\frac{ g_{t\phi} }{ \sqrt{-g_{tt}} } \right).
\label{eq:covariant-hovering_velocity}
\end{equation}

On the other hand, the energy of a particle with respect to a local observer is the time component of the four-momentum of the particle in the observer's frame of reference. It is obtained by projecting out the four-momentum of the particle on the four-velocity of the observer, i.e. $m u^\mu (u_{\mu})_{observer} = -E$.

Thus, the energy of the particle per unit mass due the frame-dragging velocity with respect to a hovering observer is
\begin{equation}
 u^\mu_{fd} u_{\mu_h} = -E_{h} = -\gamma_{fd},
\label{eq:gamma_projection}
\end{equation}
where $\gamma_{fd} = (1-v^2_{fd})^{-1/2}$ is the usual relativistic
gamma factor, $v_{fd}$ is the local velocity of the particle due
the frame-dragging, and $E_{h}$ is the relativistic energy per unit mass
of the particle relative to a stationary hovering observer. It must not be confused with the energy $E$ as seen by an observer at the infinite, at the begining of this section.

The local  velocity due the frame-dragging is then obtained from
(\ref{eq:gamma_projection}), and can be expressed as $\tanh \eta =
v_{fd}$.

Consequently, the local inertial velocity due the frame-dragging and measured by a hovering observer will be
\begin{equation}
  u^a_{fd} = \gamma_{fd}(1,~0,~0,v_{fd}).
\label{eq:fd-local_velocity}
\end{equation}

The scalar product (\ref{eq:gamma_projection}) is an invariant and
its value is independent of the coordinate system used to evaluate
it. This physical quantity in two local frames of the same event
will be connected by a Lorentz transformation between them even
though one or both of the frames may be accelerating. This follows
because the instantaneous rates of clocks and lengths of rods are
not affected by accelerations and depend only on the relative
velocities \cite{raine-thomas}.

Thereby, gathering previous results for Kerr-Newman metric we can obtain the local inertial velocity as measured by a local observer.

The angular velocity (\ref{eq:omega_formula}) on equator $\theta = \pi/2$ is
\begin{equation}
 \omega = a \left( \frac{2mr-e^2}{(r^2+a^2)^2-a^2\bar{\Delta}} \right),
\end{equation}
where positive $a$ implies positive $\omega$, so the particle acquires an angular velocity in the direction of the spin of the hole.

Therefore, as seen by a distant observer, the general four-velocities components $u^\phi_{fd}$ and $u^t_{fd}$ can be obtained from (\ref{eq:uphi_fd}) and (\ref{eq:ut_fd}), that is
\begin{equation}
 u^\mu_{fd} = \left( \frac{g_{\phi \phi} }{\bar{\Delta}} \right)^{1/2}(1,~0,~0,\omega),
\label{eq:fd_velocity}
\end{equation}

And from (\ref{eq:covariant-hovering_velocity}), (\ref{eq:gamma_projection}) and (\ref{eq:fd_velocity}), the relativistic gamma factor and local inertial frame velocity are
\begin{equation}
\begin{array}{l}
 \gamma_{fd} = \displaystyle \frac{r^2\sqrt{\bar{\Delta}} }{ \sqrt{ (\bar{\Delta}-a^2)[(r^2+a^2)^2-a^2\bar{\Delta} ]} }, \\
 v_{fd} = \displaystyle a \left( \frac{r^2+a^2-\bar{\Delta}}{r^2\sqrt{\bar{\Delta}} } \right).
\end{array}
\label{eq:KN_gamma_velocity}
\end{equation}

Finally, from (\ref{eq:fd-local_velocity}) the local  four-velocity due frame-dragging measured by a hovering observer is
\begin{equation}
u^a_{fd} =\frac{1}{\sqrt{ (\bar{\Delta}-a^2)[(r^2+a^2)^2-a^2\bar{\Delta} ]} }( r^2\sqrt{\bar{\Delta}},~0,~0,~a(r^2+a^2-\bar{\Delta})).
\label{eq:KN-fd_local_velocity}
\end{equation}

Now let $\tanh \eta =v_{fd}$, thus (\ref{eq:KN-fd_local_velocity}) can be expressed as
\begin{equation}
u^a_{fd} =(\cosh \eta,{}0,{}0,{}\sinh \eta),
\label{eq:hyperbolic-fd_local_velocity}
\end{equation}
where
\begin{equation}
  \begin{array}{l}
  \cosh \eta = \displaystyle \frac{ r^2\sqrt{\bar{\Delta}} }{ \sqrt{ (\bar{\Delta}-a^2)[(r^2+a^2)^2-a^2\bar{\Delta} ]}  }, \\
  \sinh \eta = \displaystyle \frac{ a(r^2+a^2-\bar{\Delta}) }{ \sqrt{ (\bar{\Delta}-a^2)[(r^2+a^2)^2-a^2\bar{\Delta} ]}  }.
  \end{array}
\label{eq:fd-hyperbolic_velocity}
\end{equation}

But it can be found a relative motion between the hovering observer and the local frame given by
\begin{equation}
\begin{array}{rl}
 u^a_h=  \eta^{ab}e_b{}^\mu u_{\mu_h}  & \displaystyle = \left( \sqrt{ \frac{\bar{\Delta}}{\bar{\Delta}-a^2\sin^2 \theta} },~0,~0,-\frac{a\sin \theta}{ \sqrt{{\bar{\Delta}-a^2\sin^2 \theta} } } \right) \\
  & = (\cosh \kappa,~0,~0,-\sinh \kappa),
\end{array}
\end{equation}
which implies that the hovering observer is not at rest in the local frame (\ref{eq:veirbein}).

We can remove this relative motion by a local Lorentz transformation and its inverse, that is
\begin{equation}
\begin{array}{lr}
\Lambda^a{}_b = \left(
\begin{array}{cccc}
 \cosh \kappa & 0 & 0 & \sinh \kappa  \\
 0 & 1 & 0 & 0 \\
 0 & 0 & 1 & 0  \\
 \sinh \kappa & 0 & 0 & \sinh \kappa
\end{array}
\right), &
\Lambda_a{}^b = \left(
\begin{array}{cccc}
 \cosh \kappa & 0 & 0 & -\sinh \kappa  \\
 0 & 1 & 0 & 0 \\
 0 & 0 & 1 & 0  \\
 -\sinh \kappa & 0 & 0 & \sinh \kappa
\end{array}
\right)
\end{array}
\end{equation}

Therefore, we shall consider a new vierbein $\tilde{e}_a{}^\mu = \Lambda_a{}^b e_b{}^\mu $, expressed by
\begin{equation}
\begin{array}{rl}
\tilde{e}_0{}^\mu(x) = & \left( \displaystyle \sqrt{\frac{\Sigma}{\bar{\Delta} -a^2\sin^2\theta} }, 0, 0, 0 \right), \\
\tilde{e}_1{}^\mu(x) = & \left( 0, \displaystyle \sqrt{\frac{\bar{\Delta}}{\Sigma}}, 0, 0 \right), \\
\tilde{e}_2{}^\mu(x) = & \left( 0, 0, \displaystyle \frac{1}{\sqrt{\Sigma}}, 0 \right), \\
\tilde{e}_3{}^\mu(x) = & \left( \displaystyle -\frac{ a(r^2+a^2-\bar{\Delta})\sin \theta }{\sqrt{\bar{\Delta}\Sigma}\sqrt{\bar{\Delta} -a^2\sin^2\theta}}, 0, 0, \displaystyle \frac{\sqrt{\bar{\Delta} -a^2\sin^2\theta}}{\sqrt{\bar{\Delta}\Sigma}\sin \theta}\right).
\end{array}
\label{eq:veirbein-fd}
\end{equation}

We can confirm that in the new local frame defined by (\ref{eq:veirbein-fd}), the hovering observer is at rest, i.e. $\tilde{u}^a_h=  \eta^{ab}\tilde{e}_b{}^\mu u_{\mu_h}  = (1, 0, 0, 0) $. Also it can be confirmed the local velocity of the freely falling particle $\tilde{u}^a_{fd} = \eta^{ab}\tilde{e}_b{}^\mu u_{\mu_{fd}}$ equals (\ref{eq:KN-fd_local_velocity}), which is important because we have to know the velocity due the frame-dragging  measure by a hovering observer with the right static coordinate frame. Thus, we will use this corrected veirbein for the rest of this work.

Another feature of Kerr-like spacetime is the \emph{static limit surface}. Consider a stationary particle, i.e. $r=$ constant, $\theta=$ constant and $\phi=$ constant. Thus, from spacetime metric (\ref{eq:metric})
\begin{equation}
-d\tau^2=g_{tt}dt^2.
\label{eq:static_limit_condition}
\end{equation}

Then, for $g_{tt} \geq 0$ this condition cannot be satisfied, so a massive particle cannot be stationary within the surface $g_{tt} =0$, because, as we already now, a particle will acquire four-velocity due the frame-dragging (\ref{eq:contravariant_particle_velocity}). Photons can satisfy this condition and only they can be stationary at the static limit. This is the reason why it is called static surface.

Solving the condition $g_{tt} =0$ for $r$ gives us the radius of the static limit surface
\begin{equation}
r_{st}=m+(m^2-e^2-a^2\cos^2 \theta)^{1/2}.
\label{eq:static_limit}
\end{equation}

\begin{figure}[htp]
\centering \mbox{ \includegraphics[width=4.0in,viewport=0in 0in
5.50in 3.02in,clip]{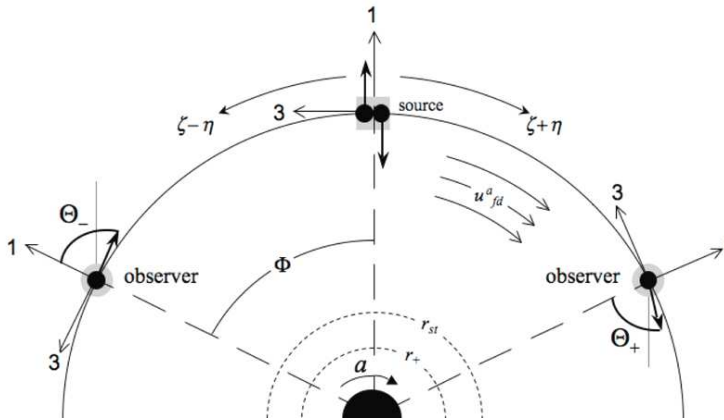} } \caption{ \small{An EPR
gedanken experiment in the Kerr-Newman spacetime with an angular
momentum parameter $a$. Two hovering observers (indicated by gray
circles) and a static EPR source (gray square) are located at
$\phi=\pm\Phi$ and 0, respectively. Both entangled particles
experiment frame-dragging $u^a_{fd}$ and leave source with a local
velocity $v=\tanh \xi =\tanh(\zeta \pm \eta)$ respect the hovering
observers, which plus sign for traveling on favor the rotation of
black hole and minus for the opposite direction.} } \label{fig:KNgedankenlab}
\end{figure}

This radius is showed in Fig.~\ref{fig:KNgedankenlab} as $r_{st}$ and is above the horizon $r_+$ as we can see. It is important to emphasize that the static limit surface is not a horizon \cite{raine-thomas}. Later, we shall see why this is no a horizon for spin precession angle, but a limit for keeping the perfect anti-correlation.

\section{Spin precession in a Kerr-Newman black hole}

\label{precession}

Now we consider massive particles with spin-1/2
in a Kerr-Newman black hole moving in a circular motion with radius
$r$ on the equatorial plane $\theta = \pi/2$. In spherical
coordinates the relevant velocity vector has two components, the
temporal and the $\phi$-coordinate at constant radius. The velocity
vector field in Minkowski's flat-space determines the motion by the
proper-velocity with $v = \tanh \xi = \sqrt{1-1/\gamma^2}$, where
$\gamma = (1-v^2)^{-1/2}$.

Following \cite{T-U} we will use the local vector
velocity $u^a = (\cosh \xi,{} 0,{} 0,{}\sinh \xi)$. Any local vector can be described on a general
reference frame through a vierbein transformation. Local velocity then
transforms as $u^{\mu}=\tilde{e}_a{}^{\mu} u^a$. Then a general
contravariant four-vector velocity is obtained as
\begin{equation}
\begin{array}{l}
    u^t = \displaystyle \frac{r}{\sqrt{\bar{\Delta}-a^2}}\cosh \xi -\frac{a(r^2+a^2-\bar{\Delta})}{ {r\sqrt{\bar{\Delta}}}\sqrt{\bar{\Delta}-a^2} } \sinh \xi, \\
    u^{\phi} = \displaystyle \frac{ \sqrt{\bar{\Delta}-a^2} }{r\sqrt{\bar{\Delta}}}\sinh \xi,
\end{array}
\label{eq:general_velocity}
\end{equation}
and the covariant vector can be obtained by lowering indices of
contravariant velocity vector  by $u_\mu = g_{\mu \nu}u^\nu$. These
velocities satisfy the normalization condition $u^{\mu}u_{\mu}=-1$
which ensure that any material particle travels with velocity lower
than speed of light\footnote{By the relativistic addition of velocities, the frame-dragging velocity
will be incorporated on $u^a_\pm = (\cosh \xi_\pm,{} 0,{} 0,{}\sinh
\xi_\pm)$ in Section \ref{EPR_correlation}, with the argument $\xi$ redefined
by $\xi_\pm=\zeta \pm \eta$. The positive sign corresponds to a
particle co-rotating with respect to the rotation of the hole and
negative for counter-rotation.}.

In order for the particle moves in orbital motion, we must apply an
external force against the centrifugal force and the gravity. The
acceleration due to this external force is given by
\begin{equation}
a^\mu (x)= u^\nu (x) \nabla_\nu u^\mu (x).
\label{eq:acceleration}
\end{equation}

On equatorial plane the acceleration then becomes
\begin{equation}
\begin{array}{rl}
a^r = &\displaystyle \frac{1}{r^3(\bar{\Delta}-a^2)}[ -\bar{\Delta}(e^2-mr)\cosh^2 \xi + 2\sqrt{\bar{\Delta}}a(e^2-mr)\cosh \xi \sinh \xi \\
& - (a^2(\bar{\Delta}+r^2-mr)-\bar{\Delta}^2) \sinh^2 \xi ].
\end{array}
\end{equation}

When the frame-dragging velocity will incorporated into velocity (\ref{eq:general_velocity})
 in Section~\ref{EPR_correlation}, the frame-dragging (\ref{eq:KN-fd_local_velocity}) will not affect the
structure of acceleration (\ref{eq:acceleration}). This is due frame-dragging velocity $u^a_{fd}$ is independent of $t$ and $\phi$. Thus the covariant derivatives of $u^t$ and $u^\phi$ with respect to $t$ and $\phi$ are not modified.

The change in the local inertial frame consists of a boost along the
1-axis and a rotation about the 2-axis by this definition
\begin{equation}
\chi^a{}_b = -u^\nu \omega_\nu{}^a{}_b
\label{eq:change_in_local_inertial_frame},
\end{equation}
where $\omega_\nu{}^a{}_b$ are the connections one-forms which are defined as
\begin{equation}
\omega_\mu{}^a{}_b = - \tilde{e}_b{}^\nu (x) \nabla_\mu \tilde{e}^a{}_\nu(x) =
\tilde{e}^a{}_\nu (x) \nabla_\mu \tilde{e}_b{}^\nu(x).
\end{equation}

Under our particular situation, the connections of our interest shall be:
\begin{equation}
\begin{array}{l}
\omega_t{}^0{}_1 = \displaystyle -\frac{(e^2-mr) \sqrt{\bar{\Delta}}}{r^3\sqrt{\bar{\Delta}-a^2}}, \\
\omega_t{}^1{}_3 = \displaystyle \frac{a(e^2-mr)}{r^3\sqrt{\bar{\Delta}-^2} }, \\
\omega_\phi{}^0{}_1 = \displaystyle \frac{a(e^2-mr) \sqrt{\bar{\Delta}}}{r^3\sqrt{\bar{\Delta}-a^2}}, \\
\omega_\phi{}^1{}_3 = \displaystyle \frac{ a^2r^2-\bar{\Delta} r^2+ma^2r-a^2e^2 }{r^3\sqrt{\bar{\Delta}-a^2}}.
\end{array}
\end{equation}

Therefore, the relevant boosts are described by the function
\begin{equation}
\chi^0{}_1 = \frac{ e^2-mr }{ r^2(\bar{\Delta}-a^2) }(\sqrt{\bar{\Delta}}\cosh \xi -a\sinh \xi),
\end{equation}
while the rotation about the 2-axis is given by
\begin{equation}
\chi^1{}_3 =\displaystyle -\frac{ 1 }{ r^2(\bar{\Delta}-a^2) }\left( a(e^2-mr)\cosh \xi +\frac{ [a^2(\bar{\Delta}+r^2-mr)-\bar{\Delta}^2] }{ \sqrt{\bar{\Delta}} }\sinh \xi \right).
\end{equation}

Next step is to join the boost and rotation\footnote{Which represent
the change on the local inertial frame along $u^\mu(x)$. } with the
rotation of the local four-momentum on the plane traced by the
general four-vectors of velocity and acceleration. Then we can
compute the infinitesimal Lorentz transformation given by
\begin{equation}
\lambda^a{}_b (x) = -\frac{1}{m}[a^a(x)p_b(x)-p^a(x)a_b(x)] +
\chi^a{}_b,
\end{equation}
where the local four-momentum defined as $p^a(x)=(m\cosh \xi,{}0,{}0,{}m\sinh \xi)$.

The boost along the 1-axis and the rotation about the 2-axis are respectively
\begin{equation}
\begin{array}{l}
\lambda^0{}_1 =-\displaystyle \frac{\sinh \xi}{r^2\sqrt{\bar{\Delta}}(\bar{\Delta}-a^2)}[ A\sinh \xi \cosh \xi -B(\cosh^2 \xi + \sinh^2 \xi) ] , \\
\lambda^1{}_3 = \displaystyle  \frac{\cosh \xi}{r^2\sqrt{\bar{\Delta}}(\bar{\Delta}-a^2)} [ A\sinh \xi \cosh \xi -B(\cosh^2 \xi + \sinh^2 \xi) ],
\end{array}
\label{eq:Lorentz_transformations}
\end{equation}
where
\begin{equation}
\begin{array}{l}
A = \bar{\Delta}^2-(a^2+mr-e^2)\bar{\Delta}-a^2r^2+a^2mr, \\
B = a\sqrt{\bar{\Delta}}(e^2-mr).
\end{array}
\end{equation}

With these results, we can get the change of the spin by the equation
\begin{equation}
\vartheta^i{}_k(x) = \lambda^i{}_k(x) + \frac{\lambda^i{}_0(x)p_k(x) - \lambda_{k0}(x)p^i(x)}{p^0(x)+m}.
\end{equation}

For our case it becomes a rotation about the 2-axis through an angle:
\begin{equation}
\vartheta^1{}_3 =  \frac{1}{r^2\sqrt{\bar{\Delta}}(\bar{\Delta}-a^2)} [ A\sinh \xi \cosh \xi -B(\cosh^2 \xi + \sinh^2 \xi) ].
\label{eq:theta13}
\end{equation}

Then, the complete rotation matrix due the infinitesimal Lorentz transformations is given by
\begin{equation}
\vartheta^a{}_b(x)= \left(
\begin{array}{cccc}
  0&0&0&0 \\
  0&0&0&\vartheta^1{}_3 \\
  0&0&0&0 \\
  0&-\vartheta^1{}_3&0&0
\end{array}
\right).
\label{eq:matrix_rotations}
\end{equation}

In non-relativistic quantum mechanics, the only kinematic
transformations of reference frames that are allowed to consider are
translations and rotations, which are explicitly unitary. In
relativistic quantum mechanics, it must also consider Lorentz
boosts, which are explicitly non-unitary, when they are represented
by finite dimensional matrices. Regardless of this, the particle
states undergoes an effective momentum dependent local unitary
rotation under boosts governed by the \emph{little group} of Wigner
rotation for massive particles, which leaves the appropriate local
rest momentum invariant. This group is SO(3) for massive particles
which is the group of ordinary rotations in 3D \cite{a-m}.

In the case of the curved spacetime, the one-particle state $
|p^a(x) , \sigma; x\rangle$ transforms under a local Lorentz
transformation as
\begin{equation}
U(\Lambda (x)) |p^a(x) , \sigma; x\rangle = \sum_{\sigma '}
D_{\sigma '\sigma}^{(1/2)}( W(x) ) |\Lambda p^a(x) , \sigma';
x\rangle, \label{eq:unitary_trans}
\end{equation}
where $W^a{}_b(x) \equiv W^a{}_b(\Lambda (x),p(x))$ is the local
Wigner rotation \cite{T-U} and $\sigma$ represents the spin state.

If a particle moves along a path $x^\mu(\tau)$ from
$x_i^\mu(\tau_i)$ to $x_f^\mu(\tau_f)$, the iteration of the
infinitesimal transformation for a finite proper time gives the
corresponding finite Wigner rotation
\begin{equation}
\begin{array}{rl}
W^a{}_b(x_f,x_i)= & \displaystyle \lim_{N \to \infty} \prod_{k=0}^N \left[ \delta^a{}_b+\vartheta^a{}_b(x_{(k)})\frac{h}{N} \right] \\
= & \displaystyle T \exp{\left( \int_{\tau_i}^{\tau_f}
\vartheta^a{}_b (x(\tau)) d\tau \right)},
\end{array}
\end{equation}
as proved in \cite{T-U}. Then a total argument $\Phi$ is
completed by integrating out $\delta \phi= u^\phi d\tau$, and the
operator $T$ is not needed because $\vartheta^a{}_b$ is constant
during the motion\footnote{The Kerr metric is the unique stationary
axial-symmetric vacuum solution as the Carter-Robinson theorem asserts
\cite{raine-thomas} and then $\vartheta^a{}_b$ is independent of the
time coordinate. }. Therefore, the velocity $u^\phi$ represents a
trivial rotation about the 2-axis
\begin{equation}
u^\phi \equiv \varphi^1{}_3 = -\varphi^3{}_1 = \displaystyle \frac{ \sqrt{\bar{\Delta}-a^2} }{r\sqrt{\bar{\Delta}}}\sinh \xi,
\label{eq:trivial_rotation}
\end{equation}
since the curved spacetime defines the parallel transport needed to
compare local inertial frames from one point to another.

Thus, the Wigner rotation becomes a rotation about the 2-axis
\begin{equation}
  \begin{array}{rl}
    W^a{}_b(\pm \Phi,0)= &\displaystyle \exp{\left( \int_0^{\Phi} \frac{\vartheta^a{}_b(x)}{\varphi^1{}_3(x)} d\phi \right)} = \exp (\frac{\Phi}{u^\phi} \vartheta^a{}_b ) \\
    =& \left(\begin{array}{cccc}
       1&0&0&0 \\
        0&\cos \Theta &0&\pm \sin \Theta \\
        0&0&1&0 \\
        0&\mp \sin \Theta &0&\cos \Theta \end{array} \right),
\end{array}
\label{eq:wigner}
\end{equation}
where $\vartheta^a{}_b$ comes from (\ref{eq:matrix_rotations}), the
angle of rotation is given by $\Theta = \Phi \vartheta^1{}_3 /
u^\phi$ and $\vartheta^1{}_3$ for the Kerr-Newman spacetime was
given by (\ref{eq:theta13}), that is
\begin{equation}
\Theta =\frac{\Phi}{r(\bar{\Delta}-a^2)^{3/2}}[A\cosh \xi -B(\coth \xi \cosh \xi - \sinh \xi)].
\label{eq:Theta}
\end{equation}

\section{EPR correlation}

\label{EPR_correlation}

In the present work we consider two observers and an EPR
source on the equator plane $\theta = \pi /2$, at a  fixed radius
above horizon ($r > r_+$), with azimuthal angles $\pm \Phi$ for
observers and 0 for the EPR source. The observers and the EPR source
are assumed to be hovering (\ref{eq:hovering_velocity_formula}) over
the black hole in order to keep them "at rest" in the Boyer-Linquist
coordinate system ($t,~r,~\theta,~\phi$) and to use the static local
inertial frame~(\ref{eq:veirbein-fd}) to measure or prepare the spin
state. The inertial frame is defined at each instant since the
observers and EPR source are accelerated to keep staying at constant
radius, and they are not influenced by the frame-dragging.

The EPR source emits a pair of entangled particles in opposite
directions. The particles adopt a circular orbit in the co-rotating
frame of the black hole due the frame-dragging
(\ref{eq:fd_velocity}). This frame corresponds to have a zero
angular momentum observers (ZAMOs). The world line of these
observers is orthogonal to the surface of constant $t$, that is,
$dx_\mu u^\mu_{fd}=0$. They have angular velocity $\omega$ as seen by a
distant observer and the angular momentum of a particle is conserved
in its local inertial frame. We will adopt a ZAMO observer as a preliminary step before we calculate the
total local inertial velocity measured by the hovering observer. The
local inertial velocity of the particles with constant four-momenta leaving
the source by EPR process is $v_\mathrm{EPR}=\tanh \zeta$ from the
point of view of a ZAMO, thus
\begin{equation}
 u^a_\mathrm{EPR} = (\cosh \zeta,~0,~0,~\sinh \zeta).
\label{eq:velocity_EPR}
\end{equation}

Therefore, from the point of view of a hovering observer, the
particles will have a local velocity given by the relativistic
addition of the velocity of ZAMOs (\ref{eq:hyperbolic-fd_local_velocity})
measured by this hovering observer, plus the local velocity of the
particles measured by ZAMOs (\ref{eq:velocity_EPR}), that is $\tanh
\xi = \tanh (\zeta \pm \eta)$, where $\xi$ comes from Section \ref{precession}.

Once the particles leave the EPR source, one travels in direction of
rotation of the black hole, and the other one travels in the
opposite direction. Then, the final constant four-momenta is given
by $p^a_\pm=(m\cosh(\zeta \pm \eta),~0,~0,~\pm m\sinh(\zeta\pm
\eta))$ measured by a hovering observer.

This incorporation of velocity due the frame-dragging redefines (\ref{eq:general_velocity}) and all calculations of Section~\ref{precession} are affected in cascade. But the general structure are not modified and neither appear new terms, as previously mentioned for the acceleration computation.

Now, the spin-singlet state is defined by
\begin{equation}
 |\psi \rangle = \frac{1}{\sqrt{2}}[ |p_+^a , \uparrow; 0 \rangle|p_-^a , \downarrow; 0 \rangle - |p_+^a , \downarrow; 0 \rangle |p_-^a , \uparrow; 0 \rangle],
\end{equation}
where for notational simplicity it was written only the $\phi$
coordinate in the arguments. All the previous considerations are schemed in a gedanken experiment shown in Fig.~\ref{fig:KNgedankenlab}.

After a proper time $\Phi/ u^\phi_\pm$, each particle reaches the
corresponding observer. The Wigner rotation~(\ref{eq:wigner})
becomes
\begin{equation}
    W^a{}_b(\pm \Phi,0)=
    \left(\begin{array}{cccc}
       1&0&0&0 \\
        0&\cos \Theta_\pm &0&\pm \sin \Theta_\pm \\
        0&0&1&0 \\
        0&\mp \sin \Theta_\pm &0&\cos \Theta_\pm \end{array} \right),
 \label{eq:wigner2}
\end{equation}
where the angle $\Theta_\pm$ is given by Eq.~(\ref{eq:Theta}) with $\xi$
substituted by $\xi_\pm=\zeta \pm \eta$. The sign depends if the
motion of the entangled particle is in direction (or in the opposite
sense) of the frame-dragging, that is
\begin{equation}
\Theta_\pm =\frac{\Phi}{r(\bar{\Delta}-a^2)^{3/2}}[A\cosh (\zeta \pm \eta) -B(\coth (\zeta \pm \eta) \cosh (\zeta \pm \eta) - \sinh (\zeta \pm \eta))].
\label{eq:Theta_plus&minus}
\end{equation}

The Wigner rotations are represented by using the Pauli matrix $\sigma_y$ as
\begin{equation}
D_{\sigma '\sigma}^{(1/2)}( W(\pm \Phi,0) ) =\exp \left( \mp i \frac{\sigma_y}{2}\Theta_\pm \right).
\end{equation}

Therefore, each particle state is transformed by the corresponding
Wigner rotation, and the new total state is given by $|\psi '
\rangle = W(\pm \Phi)|\psi \rangle$. Hence, in the local inertial
frame at $\phi=\Phi$ and $-\Phi$, each state particle is transformed
separately by
\begin{eqnarray}
|p_\pm^a , \uparrow; \pm \Phi \rangle' &=& \cos \frac{\Theta_\pm}{2} |p_\pm^a , \uparrow; \pm \Phi \rangle \pm \sin \frac{\Theta_\pm}{2} |p_\pm^a , \downarrow; \pm \Phi \rangle, \\
|p_\pm^a , \downarrow; \pm \Phi \rangle' &=& \mp \sin
\frac{\Theta_\pm}{2} |p_\pm^a , \uparrow; \pm \Phi \rangle + \cos
\frac{\Theta_\pm}{2} |p_\pm^a , \downarrow; \pm \Phi \rangle,
\end{eqnarray}
and the entangled state is described by
\begin{equation}
\begin{array}{c}
| \psi \rangle' =\displaystyle \frac{1}{\sqrt{2} }\left[\cos \left( \frac{\Theta_+ + \Theta_-}{2}\right) ( |p_+^a , \uparrow; \Phi \rangle |p_-^a , \downarrow; -\Phi \rangle - |p_+^a , \downarrow; \Phi \rangle |p_-^a , \uparrow; -\Phi \rangle) \right. \\
\left.+\displaystyle \sin \left(\frac{\Theta_+ + \Theta_-}{2}\right) ( |p_+^a , \uparrow; \Phi \rangle |p_-^a , \uparrow;
\Phi -\rangle + |p_+^a , \downarrow; \Phi \rangle |p_-^a ,
\downarrow; -\Phi \rangle ) \right].
\end{array}
\end{equation}
This result includes the trivial rotation of the local inertial
frames $\pm \Phi$, and can be eliminated by rotating the basis at
$\phi =\pm \Phi$ about the 2-axis through the angles $\mp \Phi$,
respectively, that is
\begin{eqnarray}
|p_\pm^a , \uparrow; \pm \Phi \rangle'' &=& \cos \frac{\Phi}{2} |p_\pm^a , \uparrow; \pm \Phi \rangle \pm \sin \frac{\Phi}{2} |p_\pm^a , \downarrow; \pm \Phi \rangle, \\
|p_\pm^a , \downarrow; \pm \Phi \rangle'' &=& \mp \sin
\frac{\Phi}{2} |p_\pm^a , \uparrow; \pm \Phi \rangle + \cos
\frac{\Phi}{2} |p_\pm^a , \downarrow; \pm \Phi \rangle.
\end{eqnarray}

With this basis, the state is written as
\begin{equation}
\begin{array}{c}
| \psi \rangle'' =\displaystyle \frac{1}{\sqrt{2} }[\cos \Delta ( |p_+^a , \uparrow; \Phi \rangle' |p_-^a , \downarrow; -\Phi \rangle' - |p_+^a , \downarrow; \Phi \rangle' |p_-^a , \uparrow; -\Phi \rangle') \\
+ \sin \Delta ( |p_+^a , \uparrow; \Phi \rangle' |p_-^a , \uparrow;
\Phi -\rangle' + |p_+^a , \downarrow; \Phi \rangle' |p_-^a ,
\downarrow; -\Phi \rangle' )],
\end{array}
\label{eq:final_state}
\end{equation}
where $\Delta = (\Theta_+ + \Theta_-)/2 - \Phi$. After some
computations, the angle $\Delta$ is simplified to
\begin{equation}
 \Delta= \Phi \left\{ \frac{\cosh \eta}{r(\bar{\Delta}-a^2)^{3/2}} \left[ A\cosh \zeta - B\sinh \zeta \left( \frac{2\cosh^2 \zeta-\cosh^2\eta-\sinh^2\eta}{\cosh^2\zeta-\cosh^2\eta} \right) \right] - 1 \right\}.
\label{eq:KN_delta}
\end{equation}
This is precisely the general relativistic effect that deteriorates the
perfect anti-correlation in the directions that would be the same as
each other if the spacetime were flat and the velocities of the
particles would not be relativistic. The spin-singlet state is mixed
with the spin-triplet state. This is because while the spin-singlet
state is invariant under spatial rotations, it is not invariant
under Lorentz transformations (\ref{eq:Lorentz_transformations}).

This deterioration of the perfect anti-correlation is consequence of
the manifest difference between the rotation matrix element
$\vartheta^1{}_3$ and trivial rotation $\varphi^1{}_3$. It is
important to note that the entanglement is still invariant under
local unitary operations, and then it does not mean to put away the
nonlocal correlation. Because the relativistic effect arises from
acceleration and gravity, the perfect anti-correlation can still be
employed for quantum communication, by rotating the direction of
measurement about the 2-axis through the angles $\mp \Theta$ in the
local inertial frames of the hovering observers. The parallel
transport in general relativity
(\ref{eq:change_in_local_inertial_frame}) does not give the
directions that maintain the perfect anti-correlation, because the
rotation matrix elements (\ref{eq:matrix_rotations}) and the
components of the change in local inertial frame
(\ref{eq:change_in_local_inertial_frame}) are not equal.

\section{Kerr-Newman spin precession results}

\label{results} 
\begin{figure}[htp]
\centering
\includegraphics[width=3in,viewport=0in 0in 6.49in 5.53in,clip]{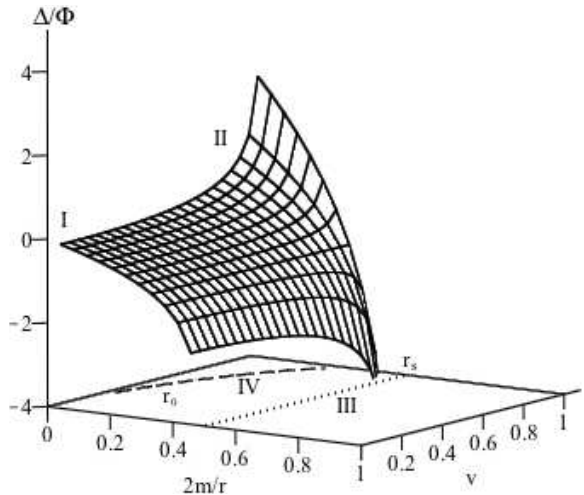}
\caption{\small{The angle $\Delta/\Phi$ for a Schwarzschild black hole as function of $2m/r$ and $v$, which is asymptotic to the event horizon $r_\mathrm{s}=2m$, indicated by a dotted line. Dashed line depicted the path $r=r_0$ which the spin precession $\Delta$ vanishes. }}
\label{fig:Schwarzschild}
\end{figure}

As Terashima and Ueda showed \cite{T-U} for a
Schwarzschild black hole, the acceleration and gravity deteriorate
the EPR correlation for particles in a circular motion in equatorial
plane. We synthesized their results in four important regions
relative the black hole plotted in Fig.~\ref{fig:Schwarzschild}.

\newcounter{region}
\begin{list}{\upshape Region \Roman{region}:}
 {\usecounter{region}
 \setlength{\labelwidth}{2cm}\setlength{\leftmargin}{2.6cm}
 \setlength{\labelsep}{0.5cm}\setlength{\rightmargin}{0cm}
 \setlength{\parsep}{0.5ex plus0.2ex minus0.1ex}
 \setlength{\itemsep}{0ex plus0.2ex} }
  \item $r \to \infty$, $v \to 0$, or far away the black hole (no gravitational effects) and static particles. This region corresponds to the non-relativistic limit, where there are no corrections to quantum mechanics and where EPR proposed their \emph{gedanken} experiment \cite{EPR}. The precession angle vanishes ($\Delta=0$) and we get the maximal violation of Bell's inequality.
  \item  $r \to \infty$, $v \to 1$, it is still far away from the black hole but it must be added relativistic corrections, which were also studied by Terashima and Ueda on \cite{t-u_sr1}. The angle $\Delta$ is positive and becomes infinite. It is no possible to maintain perfect anti-correlation and the particles cannot be used to quantum communication.
  \item $r \to r_\mathrm{s}$, where $r_\mathrm{s}=2m$ is the Schwarzschild radius (event horizon). Independently of local inertial velocity of the particles, the precession angle becomes infinite ($\Delta \to -\infty$). The static observers cannot extract the EPR correlation from circularly moving particles unless they have infinite accuracy as to their own positions. To exploit the EPR correlation on and beyond the horizon, the observers must choose a four-velocity and a non-singular vierbein at the horizon, and thus the observers must fall into the black hole together with the particles \cite{T-U}.
  \item Although acceleration and gravity deteriorate the EPR correlation as Terashima and Ueda showed, it is still possible to find a combination of local inertial velocity and position respect to the black hole that keeps the perfect anti-correlation. They defined a path where at radius $r=r_0$ the angle $\Delta$ vanishes. We will identify this path as an additional region and it is between the other three regions.
\end{list}

Between these regions one can find values of the angle (positive or
negative) $\Delta$, which deteriorates the perfect anti-correlation
in the directions that would be the same as each other if the
spacetime were flat.

We shall compere the Schwarzschild and Kerr-Newman spacetimes and we will find interesting differences. Also, with the results of Section~\ref{EPR_correlation} we will analyze the
influence of each parameter on the spin precession. There will be
remarkable differences among these regions.

The plots presented in this paper are dimensionless, where the mass
parameter is used as a reference to express the charge and angular
momentum ratio, represented by $e/m$ for electric charge, $a/m$ for
angular momentum. One of the axis plots $v=v_\mathrm{EPR}$ for the local
inertial velocity due the EPR process and $0< m/r< 1$ for distance,
with 0 corresponding to $r$ at infinite and 1 for $r=m$, which is
the smaller distance reached for extreme black holes.

\subsection{Reissner-Nordstr\"{o}m case}

This case corresponds to a Schwarzschild black hole with a
non-vanishing charge $e$.  It should be noted that, in geometric units, the charge to mass
ratio of a proton is $q/m \sim 10^{18}$, and for an electron is
$q/m \sim 10^{21}$. Since the ratio of electromagnetic to
gravitational force produced on a test body of charge $q$ and mass
$m$ by a body of charge $e$ and mass $M$ is $ \sim qe/mM$, it would
be very difficult for any astrophysical process to achieve and
maintain a charge to mass ratio greater than $ \sim 10^{-18}$, since
a body with large charge  to mass ratio would selectively attract
particles of opposite charge \cite{Wald}. Hence, on astrophysical
reasonable situation it appears that $e \ll M$. We considered
anyway an arbitrary charge to illustrate the effect of this
parameter on the spin precession, but the electromagnetic interaction between
charged particles and the charged black hole was not accounted.

From the Kerr-Newman spacetime, when $a = 0$ we recover the
Reissner-Nordstr\"{o}m solutions and spin precession
(\ref{eq:KN_delta}) reduces to
\begin{equation}
\Delta_\mathrm{RN} = \Phi \left[ \frac{r^2-3mr+2e^2}{r\sqrt{r^2-2mr+e^2} }\cosh\zeta -1 \right].
\label{eq:RN_delta}
\end{equation}
This expression has physical meaning when $e \leq m$, which is a
direct consequence from the event horizon (\ref{eq:horizons}) for
black holes.

\begin{figure}[htp]
\centering
\parbox{3in}{ \centering \includegraphics[width=3in,viewport=0in 0in 6.37in 6.85in,clip]{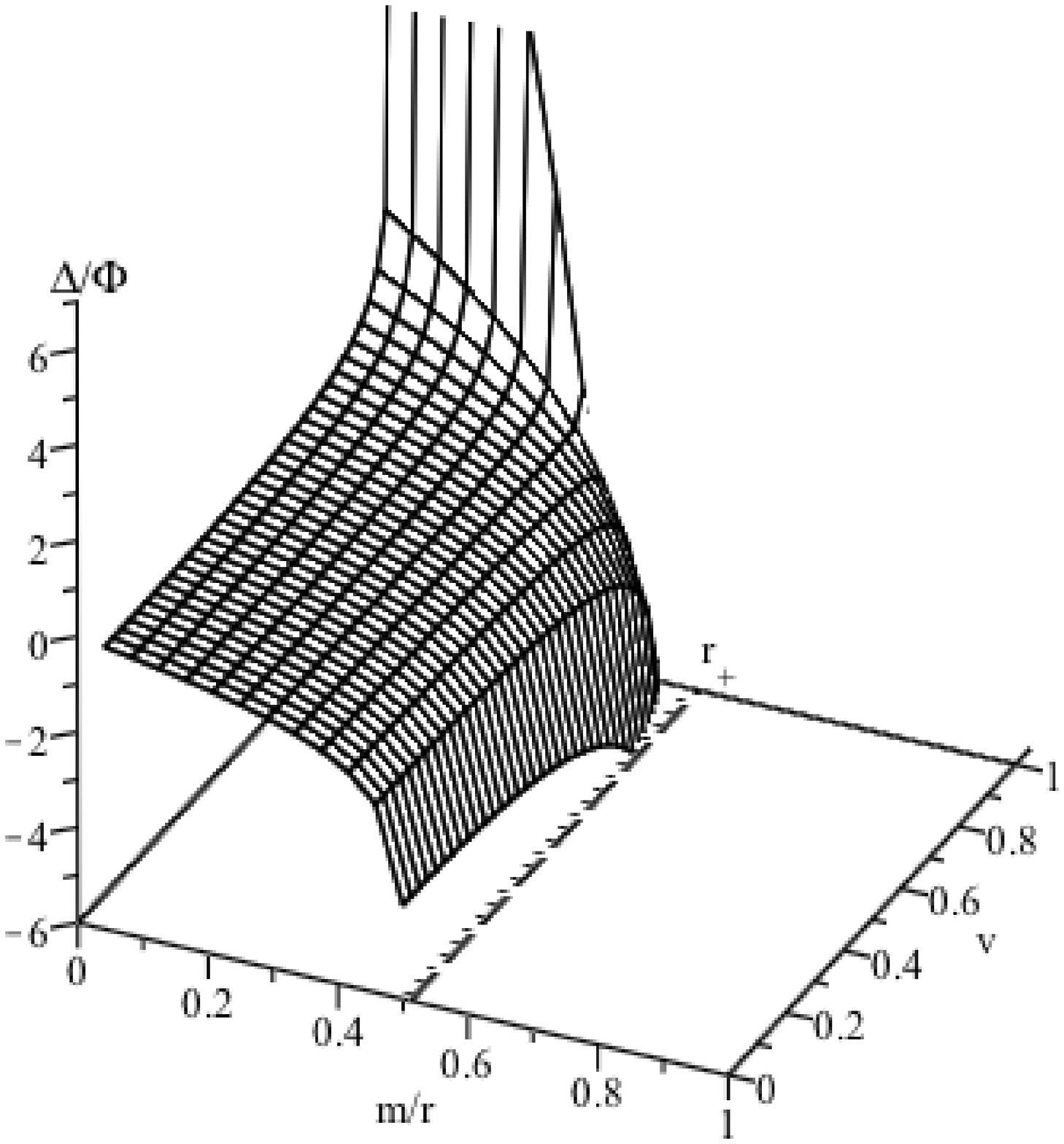}
\small{a) $e=0.3m$} }
\parbox{3in}{\centering \includegraphics[width=3in,viewport=0in 0in 6.37in 6.85in,clip]{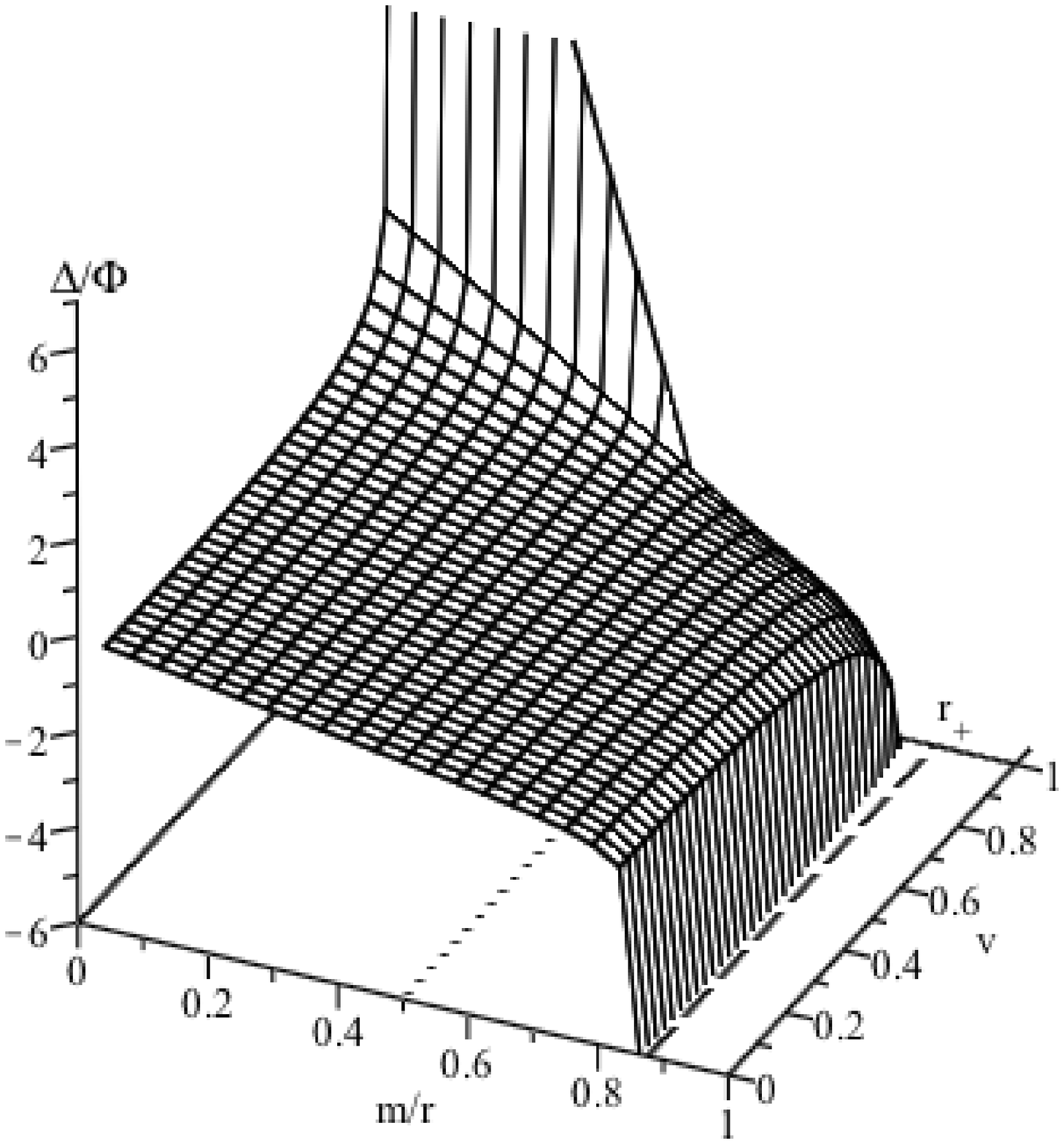}
\small{b) $e=0.99m$} }
\caption{
\small{The precession angle $\Delta / \Phi$ for a Reissner-Nordstr\"{o}m black hole for two values of charge $e$. They are asymptotic to the horizon $r=r_+$ (dashed line), which is below the Schwarzschild radius $r_\mathrm{s}=2m$ (dotted line).}}
\label{fig:RN_precession}
\end{figure}

The angle $\Delta$ on (\ref{eq:RN_delta}) is plotted in
Fig.~\ref{fig:RN_precession} as function of the distance and local
velocity $v=v_\mathrm{EPR}$. When $m/r \to 0$ the experiment is
placed far away of the black hole ($r \to \infty$), and $m/r=1$
corresponds to the limit of distance that we can reach for
calculations for an extreme black hole with charge $e=m$. When $v=0$
the particles don't have local velocity due the EPR process and for
$v \to 1$ they are ultra-relativistic particles. The precession
angle was plotted independently from the observer position angle
$\Phi$. For $e =0$ we recover all results of the spin precession in
a Schwarzschild black hole and the horizon is at $r=2m$.

The plots are quite similar as in the Schwarzschild case (compare with
Fig.~\ref{fig:Schwarzschild}). Analogous and interesting effects of
spin precession can be compared with \cite{T-U} using the previous
reviewed regions:
\newcounter{region2}
\begin{list}{\upshape Region \Roman{region2}:}
 {\usecounter{region2}
 \setlength{\labelwidth}{2cm}\setlength{\leftmargin}{2.6cm}
 \setlength{\labelsep}{0.5cm}\setlength{\rightmargin}{0cm}
 \setlength{\parsep}{0.5ex plus0.2ex minus0.1ex}
 \setlength{\itemsep}{0ex plus0.2ex} }
\item The situation is identical to the Schwarzschild black hole. The spacetime is Minkowskian and $\Delta \to 0$.
\item Far away from the horizon $r_+$ with $v \neq 0$ we recover the spin precession found in special relativity and the plot is asymptotic (see Fig.~\ref{fig:RN_precession}), i.e. $\lim_{r \to \infty}\Delta = \cosh \zeta-1$ in agreement to \cite{t-u_sr1}.
\item A new effect occurs near the black hole horizon. This effect corresponds to a shifting on horizon compared with the Schwarzschild case, from $r=2m$ to $r=r_+= m+\sqrt{m^2-e^2}$. As the charge $e$ is increased, we reach values of $r$ below the Schwarzschild horizon, it means that we can calculate values of $\Delta$ at $r=r_+ < 2m$ (see Fig.~\ref{fig:KNgedankenlab} where $r_{st}=2m$). The lowest value of $r$ that we can reach is when $e=m$ for a extreme black hole, whose horizon is at $r=r_+=m$. These values of $r$ are not allowed for the Schwarzschild case. From Fig.~\ref{fig:RN_precession} we see how the horizon is shifted as the charge is increased. $\Delta \to -\infty$ as the horizon is reached, no matter the velocity of the particles considered. EPR correlation then is totally lost. The same behavior of $\Delta$ was present in Schwarzschild radius in Ref.~\cite{T-U}.

The divergence of the spin precession originates from the fact the vierbein (\ref{eq:veirbein-fd}) and the four-velocity (\ref{eq:general_velocity}) become singular at the horizon $r_+$. These singularities are connected with the breakdown of the coordinate system ($t$, $r$, $\theta$, $\phi$).

\begin{figure}[htp]
\centering
\mbox{\includegraphics[width=2.5in,viewport=0in 0in 7.38in 7.41in,clip]{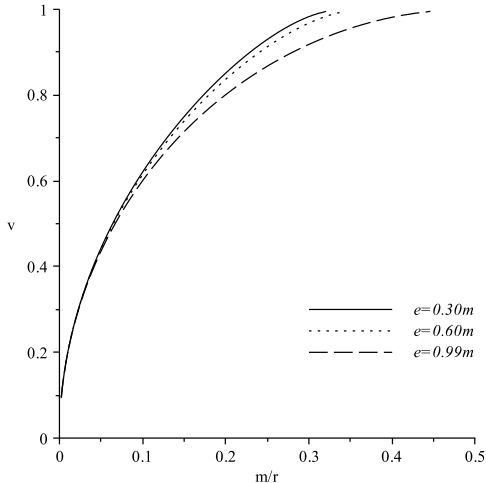}}
\caption{\small{Parametric plot of position $m/r$ and local inertial velocity $v$ for path  $r_0$  that keep a perfect anti-correlation ($\Delta=0$) for a Reissner-Nordstr\"{o}m black hole. }}
\label{fig:RN_perfect_anticorrelation}
\end{figure}

\item It is still possible to keep circular orbits in the path $r=r_0$, with perfect anti-correlation $\Delta=0$. Thus, for a particular position, the local inertial velocity of particles $v_\mathrm{EPR}$ must be settled at the beginning from the source,  In Fig.~\ref{fig:RN_perfect_anticorrelation} $r_0$ is plotted for three suitable values of charge $e$ in function of position $m/r$ and local inertial velocity $v$. We can see that  for large distance ($m/r \to 0$) we can keep the perfect anti-correlation with low values of $v$. Meanwhile the horizon is reached, we must increase the local velocity of the particles to keep the perfect anti-correlation.

As in the Schwarzschild case, near the horizon there is not null
precession angle ($\Delta \neq 0$), independently of the velocity of
the particle. Then it is not possible to have a perfectly
anti-correlated orbits. In Fig.~\ref{fig:RN_perfect_anticorrelation}
the limit circular orbits correspond to the point where $r_0$ ends
on the top of the figure. For large values of $e$, we can have
perfect anti-correlated orbits closer to the horizon, but there is no possible to find a $r_0$ below the Schwarzschild radius.

\end{list}

\subsection{Kerr case}
\label{Kerr_result}

Now we consider a rotating black hole without charge. It corresponds
to the Kerr spacetime. The spin precession has the same form of
(\ref{eq:KN_delta}), but with different coefficients $A$ and $B$, that is, after setting $e=0$ we get
\begin{equation}
\Delta_\mathrm{K} = \Phi \left\{ \frac{\cosh \eta}{r(r^2-2mr)^{3/2}} \left[ A_\mathrm{K} \cosh \zeta - B_\mathrm{K} \sinh \zeta \left( \frac{2\cosh^2 \zeta-\cosh^2\eta-\sinh^2\eta}{\cosh^2\zeta-\cosh^2\eta} \right) \right] - 1 \right\},
\label{eq:Kerr_delta}
\end{equation}
where
\begin{equation}
\begin{array}{l}
A_\mathrm{K} = r^4-5mr^3+6m^2r^2-2a^2mr,\\
B_\mathrm{K} =  -amr\sqrt{r^2-2mr+a^2}.
\end{array}
\end{equation}

\begin{figure}[htp]
\centering
\parbox{3in}{ \centering \includegraphics[width=3in,viewport=0in 0in 6.5in 6.7in,clip]{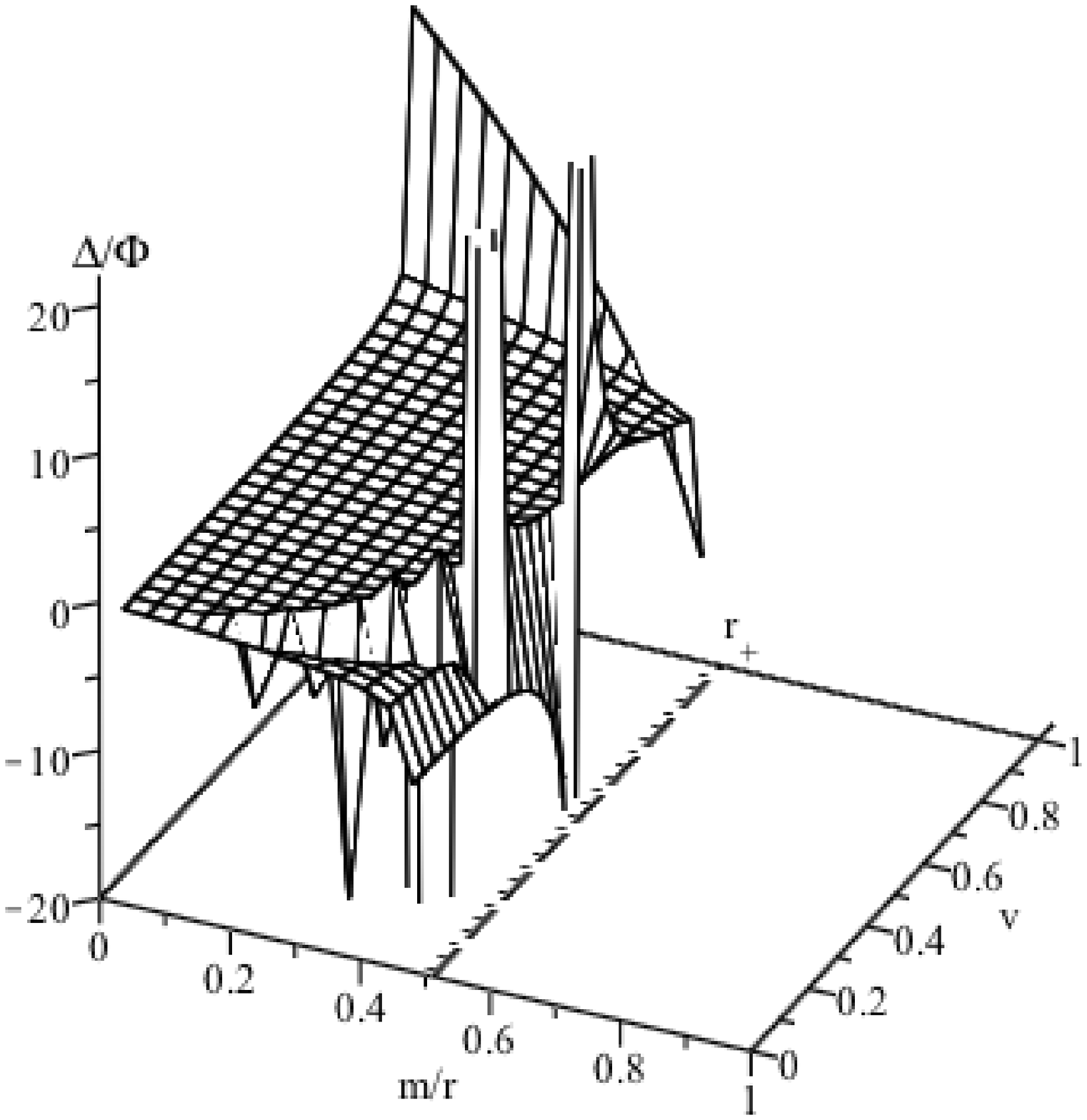}
\small{a) $a=0.3m$} }
\parbox{3in}{\centering \includegraphics[width=3in,viewport=0in 0in 6.5in 6.71in,clip]{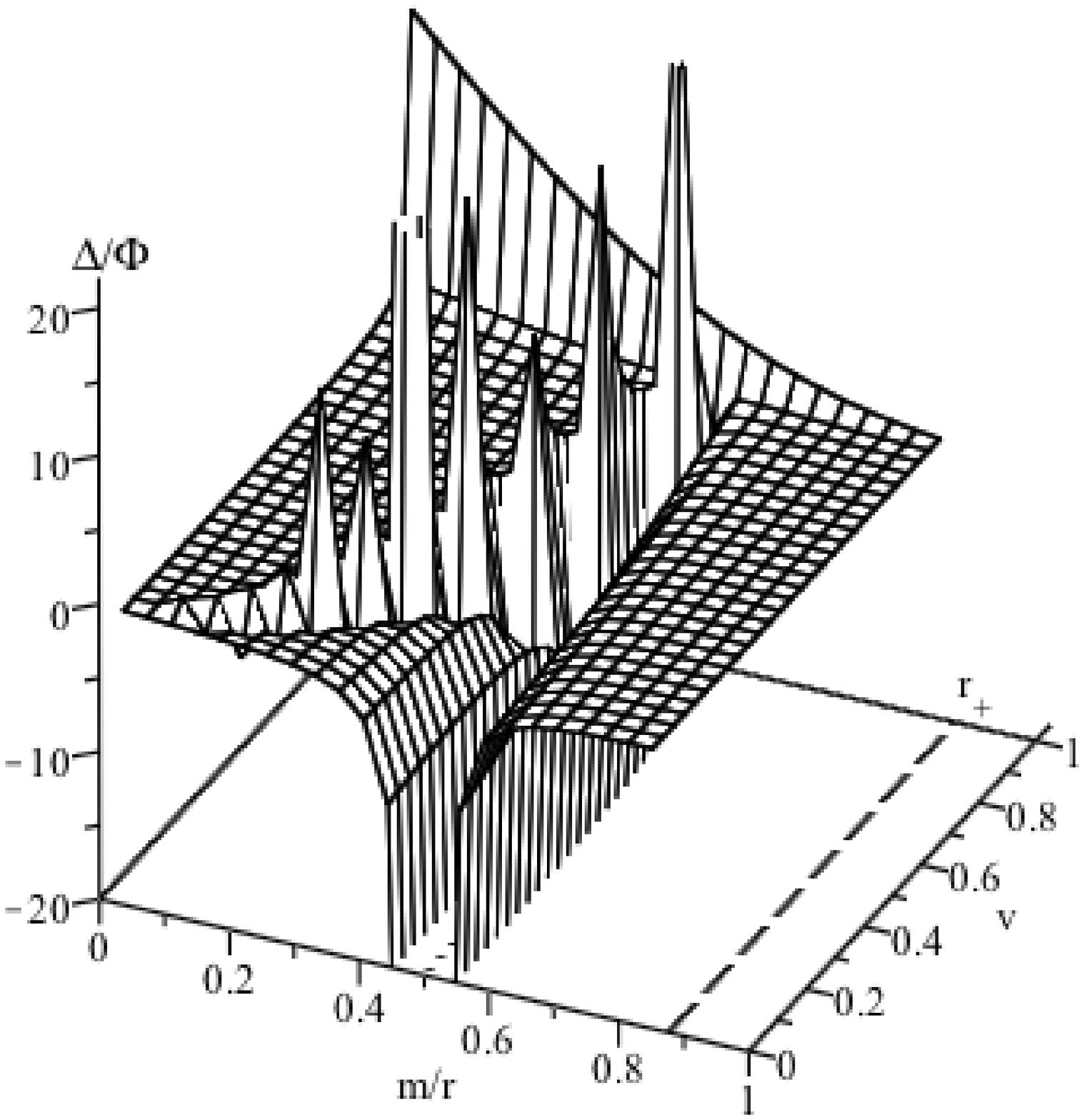}
\small{b) $a=0.99m$} }
\caption{\small{The precession angle $\Delta / \Phi$ for a Kerr black hole for two values of angular momentum parameter $a$. They are asymptotic to the static limit $r_{st}=2m$ and along a path $v=v_{fd}$. The peaks represent an asymptotic infinite wall.}}
\label{fig:Kerr_precession}
\end{figure}

The precession angle is plotted in Fig.~\ref{fig:Kerr_precession} for two values of angular momentum parameter $a$, as function of distance and local velocity $v=v_\mathrm{EPR}$. The distance is parameterized by $m/r$ which means the experiment is placed at infinite when $m/r \to 0$, and $m/r=1$ correspond to a '`extreme" black hole, i.e. $a=m$. When $v=0$ the particles do not have a local velocity due the EPR process and for $v \to 1$ they are ultra-relativistic particles. The precession angle was plotted independently from the observer position angle $\Phi$. For $a =0$ we recover all results of Schwarzschild spin precession \cite{T-U} as expected.

The plot is quite similar to Fig.~\ref{fig:Schwarzschild}, but with important differences. The effects due the spacetime metric analyzed by regions are:
\newcounter{region3}
\begin{list}{\upshape Region \Roman{region3}:}
 {\usecounter{region3}
 \setlength{\labelwidth}{2cm}\setlength{\leftmargin}{2.6cm}
 \setlength{\labelsep}{0.5cm}\setlength{\rightmargin}{0cm}
 \setlength{\parsep}{0.5ex plus0.2ex minus0.1ex}
 \setlength{\itemsep}{0ex plus0.2ex} }
\item Again the situation is identical to the Schwarzschild's black hole. The frame-dragging has no contribution because it weakens with distance. Therefore the spacetime is Minkowskian and $\Delta \to 0$ as $v=v_\mathrm{EPR} \to 0$.
\item There are no new effects. The frame-dragging has no contribution and the angle $\Delta$ is asymptotic to infinite when $v \to 1$ for ultra-relativistic particles.
\item   In the Schwarzschild and Reissner-Nordstr\"{o}m spacetime, the divergence of the spin precession ($\Delta \to -\infty$) was at the horizon. Now, the divergence is present in two locations, one of them at the static limit surface and the other one is through the path defined by $v_\mathrm{EPR} = v_{fd}$.

The first divergence in (\ref{eq:Kerr_delta}) is related to the static limit surface. As mentioned in Section \ref{frame-dragging}, any particle acquire velocity due the frame-dragging as it falls to the black hole. When this particle reaches the static limit surface at $r=2m$ for equatorial plane, its velocity tends asymptotically to speed of light. In the left part of equation (\ref{eq:Kerr_delta}) it is easy to see why precession angle diverges when distance is evaluated at $2m$. The divergence of the spin precession in the Kerr spacetime originates from the fact that the frame-dragging component (\ref{eq:KN-fd_local_velocity}) of the four-velocity (\ref{eq:general_velocity}) becomes singular at the static limit $r_{st}$. This feature contrasts with the Reissner-Nordstr\"{o}m case, where the singularities were connected with the breakdown of the coordinate system ($t$, $r$, $\theta$, $\phi$) at $r_+$.

Previously was mentioned that the static limit is not a horizon. Beyond $r_{st}$ it is still  possible to get entangled particles in circular orbits. The region between $r_+ \leqslant r < 2m$ has a similar behavior as Region I and II (see in particular Fig.~\ref{fig:Kerr_precession}~b) where is more clear this feature). Frame-dragging has no effect and the precession angle $\Delta_\mathrm{K}$ is asymptotic near the static limit at $2m$ and also for ultra relativistic particles. Near the horizon $r_+$ the function (\ref{eq:Kerr_delta}) is well defined. This is an unexpected result if we compare with Scharzschild and Reissner-Nordstr\"{o}m cases, where the horizon represents an asymptotic limit. For $r < r_+$ the coordinate system breakdowns and we are unable to find the precession angle for orbital particles.

The second divergence corresponds to a coupling between the EPR velocity and the frame-dragging. In Fig.~\ref{fig:Kerr_precession} we can see an asymptotic infinite wall in the figure. This asymptotic path happens when $\cosh^2 \zeta = \cosh^2 \eta$ in (\ref{eq:Kerr_delta}), which is easy to verify that corresponds to $v_\mathrm{EPR} = v_{fd}$.

When the velocity of the first particle equals the velocity of the frame-dragging, $\Delta_\mathrm{K}$ becomes asymptotically infinity. Physically, one particle stay static because $v_\mathrm{EPR}$ equals $v_{fd}$, meanwhile, the other particle continues his travel, as seen by the hovering observer. The static particle never reach the observer, and there are no means to know the anti-correlation between the particles.

This situation represents a particular feature for Kerr-like spacetime. In Schwarzschild, Reissner-Nordstr\"{o}m and \cite{S-A} the plots were very smooth until their functions reach their horizons. Here, the plot presents a sudden infinite wall, which follow the velocity that experience a free falling particle due the frame-dragging (see Fig.~\ref{fig:fd&anticorrelation}~a).

\begin{figure}[htp]
\centering
\parbox{2.6in}{\centering \includegraphics[width=2.5in,viewport=0in 0in 7.38in 7.41in,clip]{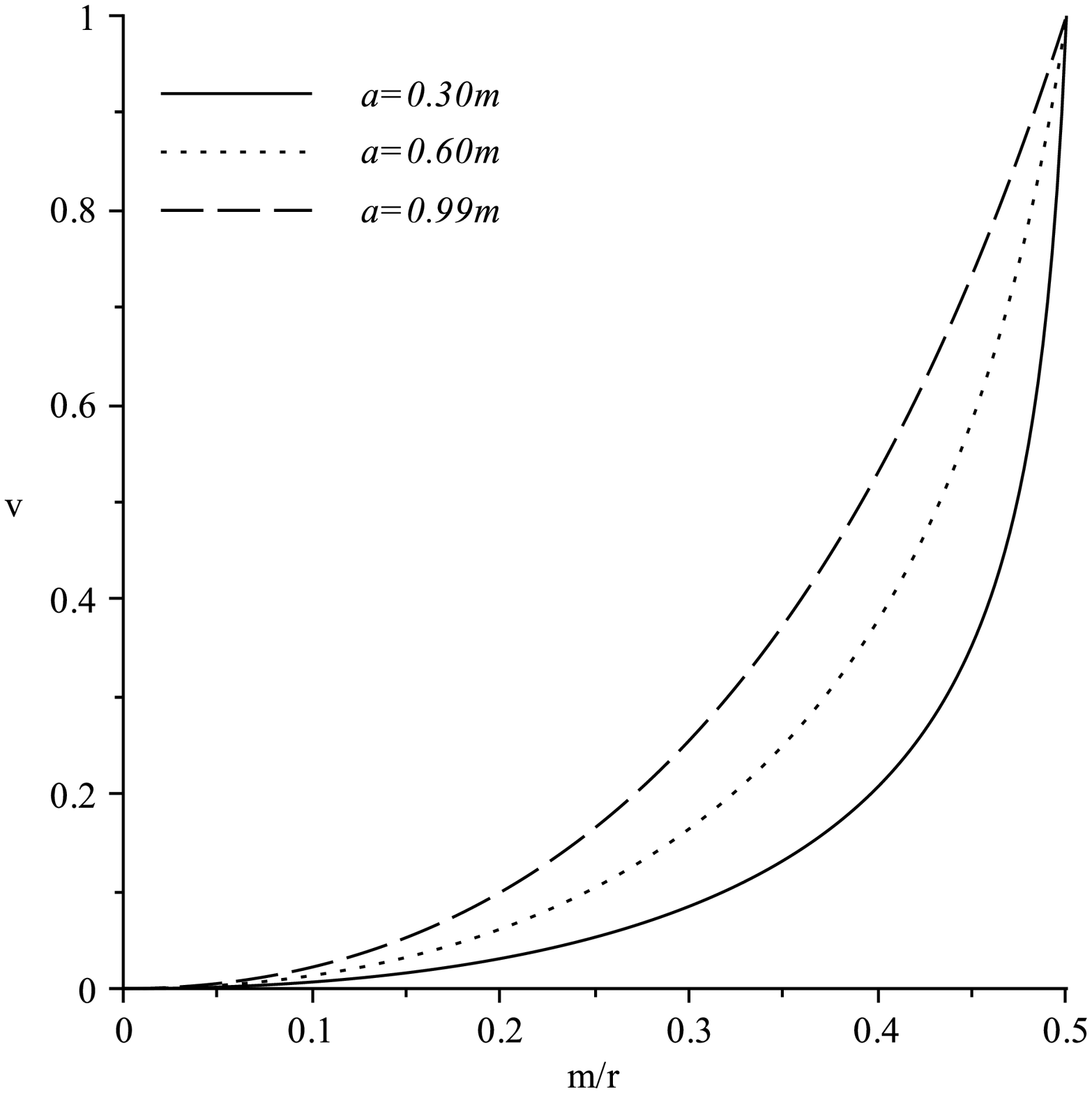} \small{a) Frame-dragging velocity, $v = v_{fd}$} }
\parbox{2.6in}{\centering \includegraphics[width=2.5in,viewport=0in 0in 7.38in 7.41in,clip]{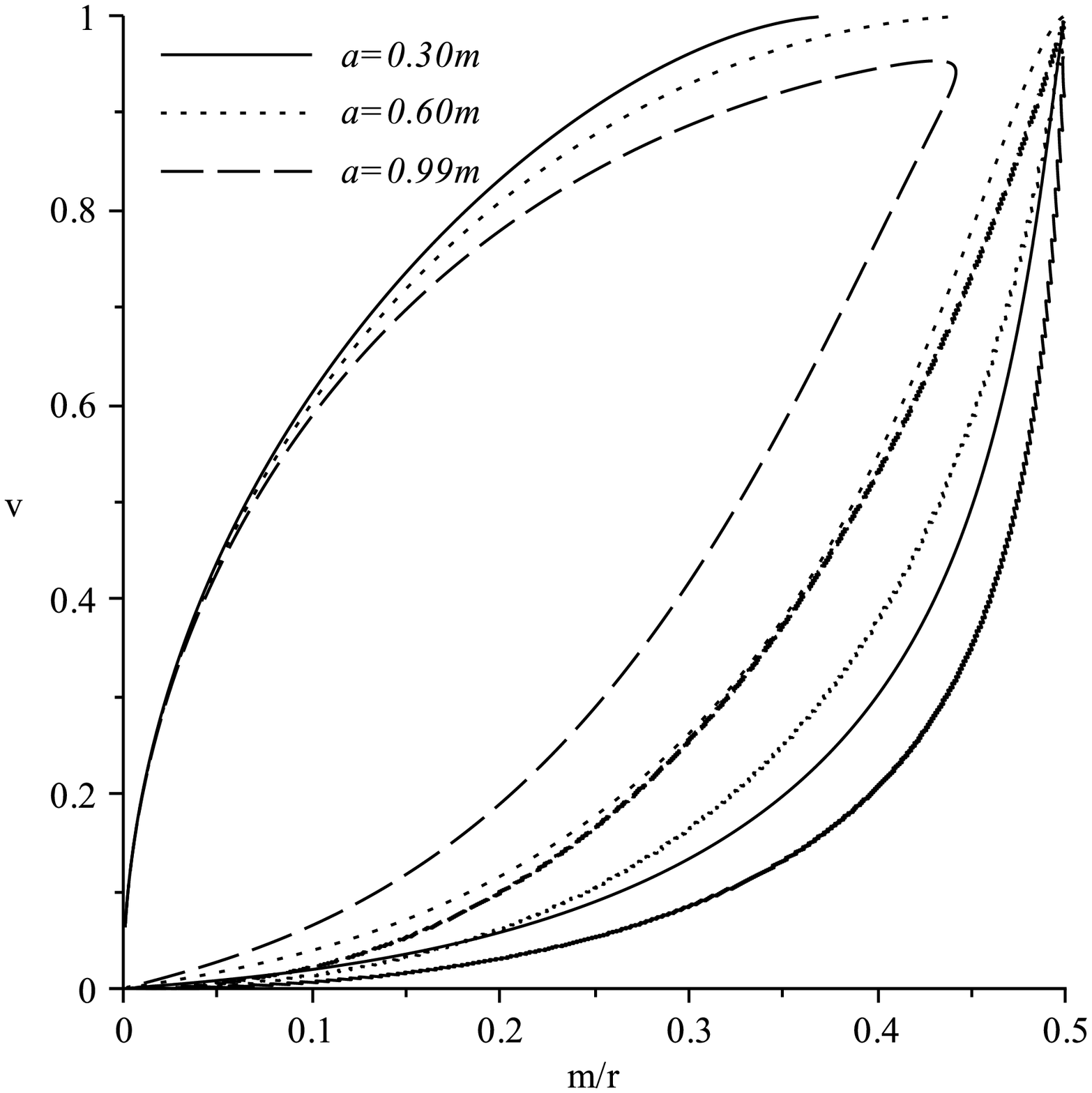} \small{b) Path $r_0$ for Kerr spacetime} }
\caption{\small{a) Local inertial velocity due the frame-dragging for three values of $a$. The infinite wall in Fig.~\ref{fig:Kerr_precession} follow the path traced by this plot when $v_\mathrm{EPR}=v_{fd}$. b) Parametric plot of position $m/r$ and local inertial velocity $v$ for path  $r_0$  that keep a perfect anti-correlation ($\Delta=0$) for a Kerr black hole.} }
\label{fig:fd&anticorrelation}
\end{figure}

\item We can see in Fig.~\ref{fig:fd&anticorrelation}~b) that away from black hole, there is a low velocity that keeps the perfect anti-correlation, as in the Schwarzschild and Reissner-Nordstr\"{o}m cases. As $r \to r_{st}$ there is a non-vanishing precession angle, independently of the velocity of the particle $v_\mathrm{EPR}$. Perfectly anti-correlated orbits cannot be kept and $r_0$ has a limit value as in the Reissner-Nordstr\"{o}m spacetime. We can see this limit value when $r_0$ ends on the right top the Fig.~\ref{fig:fd&anticorrelation}~b). Near the static limit, the contribution of the frame-dragging allows three values of $v_\mathrm{EPR}$. This new effect is not present in the Schwarzschild and Reissner-Nordstr\"{o}m cases and in the previous work \cite{S-A} neither.
When the static limit is reached, the velocity due the EPR process must be the speed of light, making impossible to keep the perfect anti-correlation.
\end{list}

\subsection{Kerr-Newman case}

We are now in position to analyze the complete Kerr-Newman spacetime
and its effects on entangled particles.  From (\ref{eq:KN_delta}) it
can be shown that Region I and II have the same behavior for a
Minkonski spacetime. This is not surprising result since as we
have seen in previous cases of this section, $a$ and $e$ weaken with
distance.

\begin{figure}[htp]
\centering
\parbox{2.6in}{ \centering \includegraphics[width=2.5in,viewport=0in 0in 6.5in 6.64in,clip]{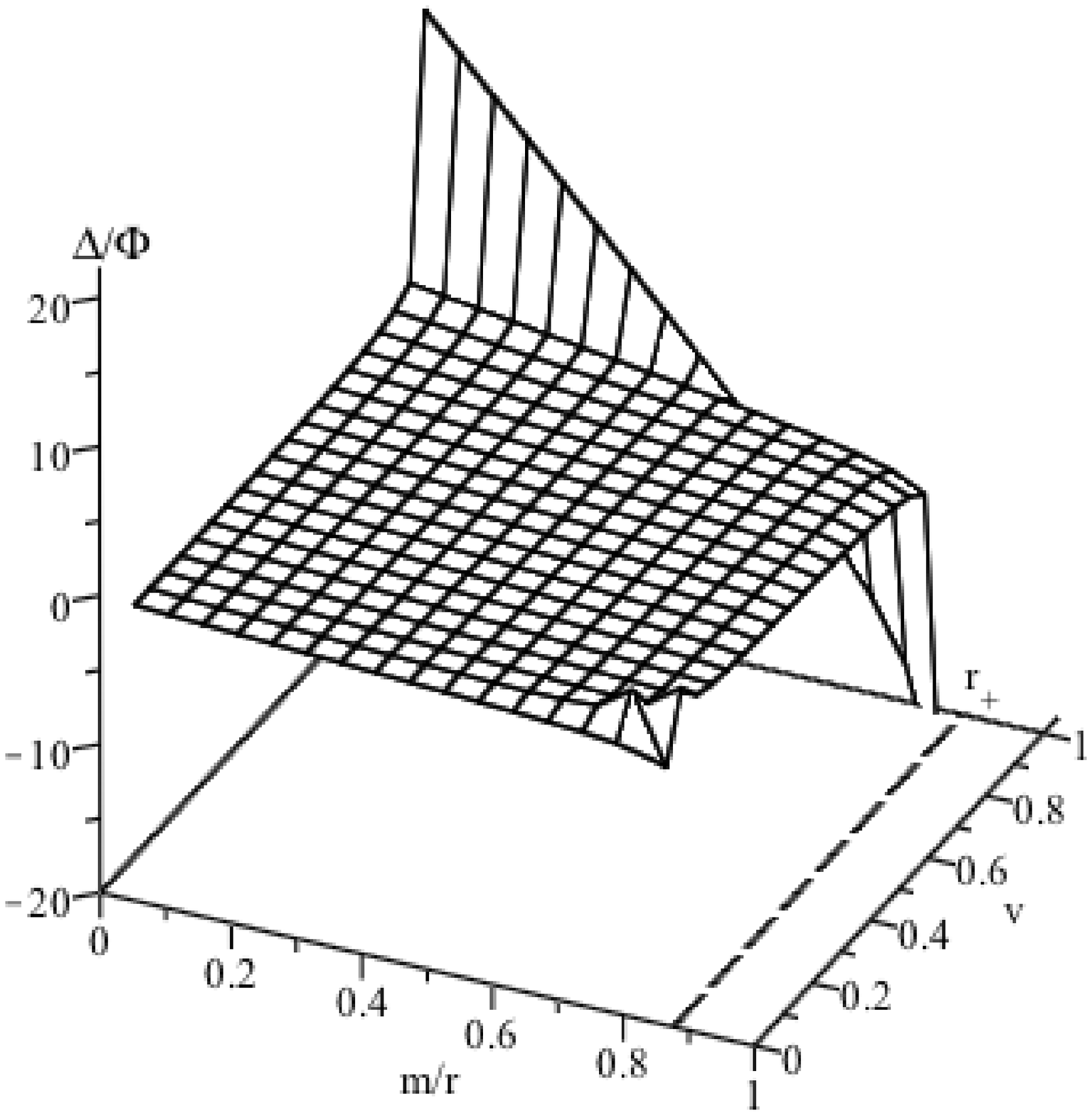}
\small{a) $a=0.01m$, $e=0.99m$ } }
\parbox{2.6in}{\centering \includegraphics[width=2.5in,viewport=0in 0in 6.54in 6.64in,clip]{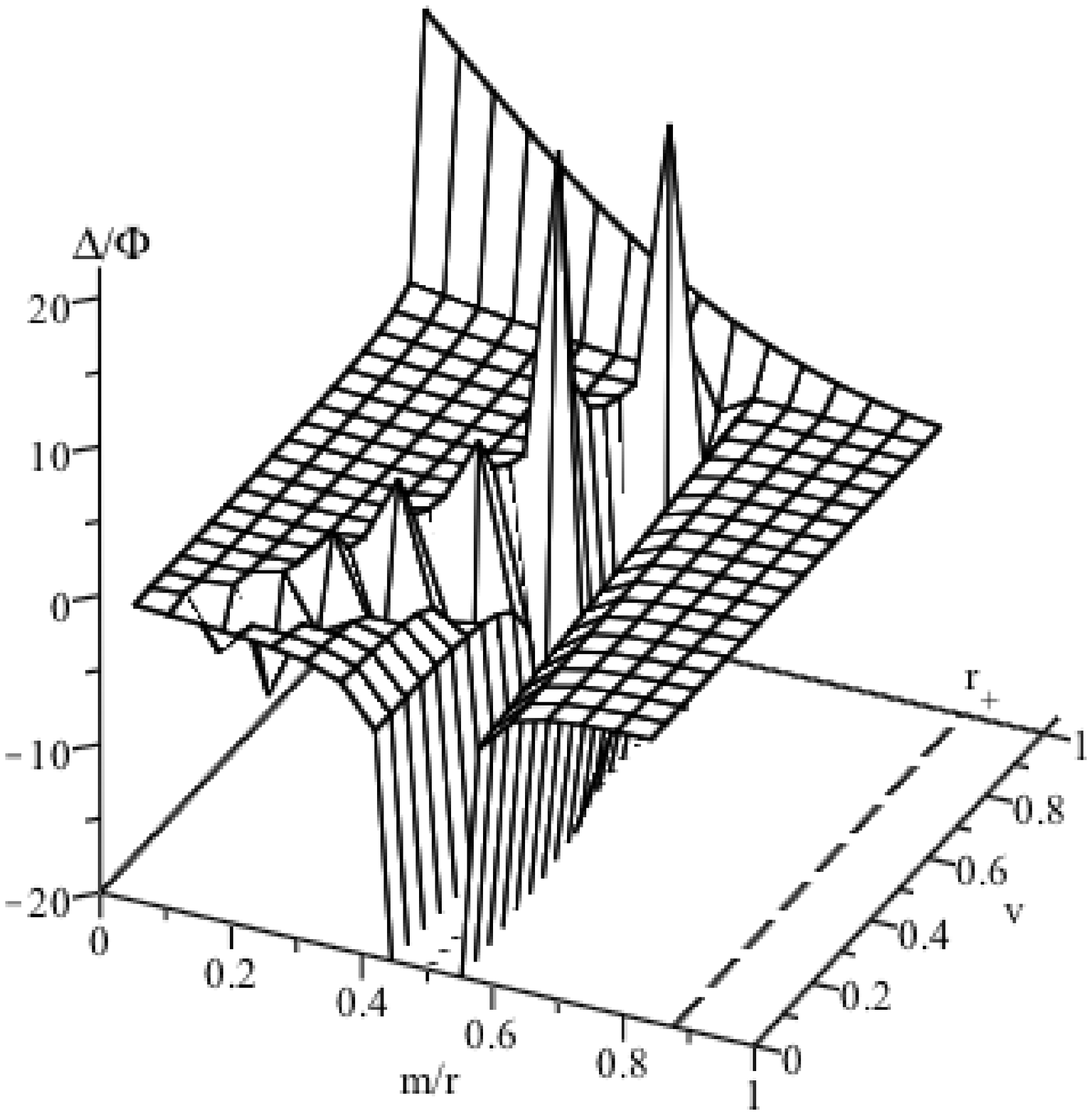}
\small{b) $a=0.99m$, $e=0.01m$ } }
\parbox{2.6in}{\centering \includegraphics[width=2.5in,viewport=0in 0in 6.54in 6.64in,clip]{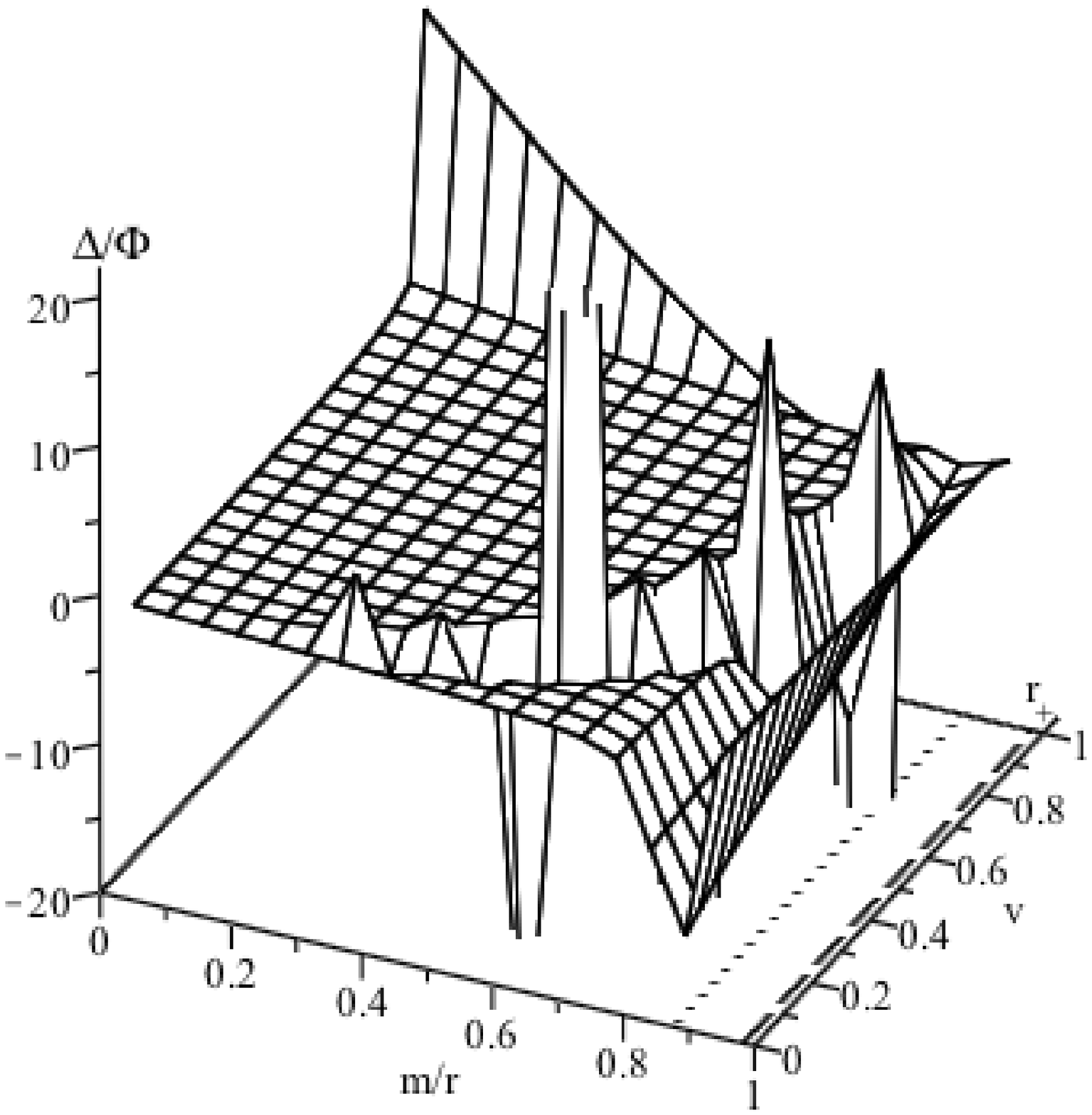}
\small{b) $a=0.14m$, $e=0.99m$ } }
\parbox{2.6in}{\centering \includegraphics[width=2.5in,viewport=0in 0in 6.54in 6.64in,clip]{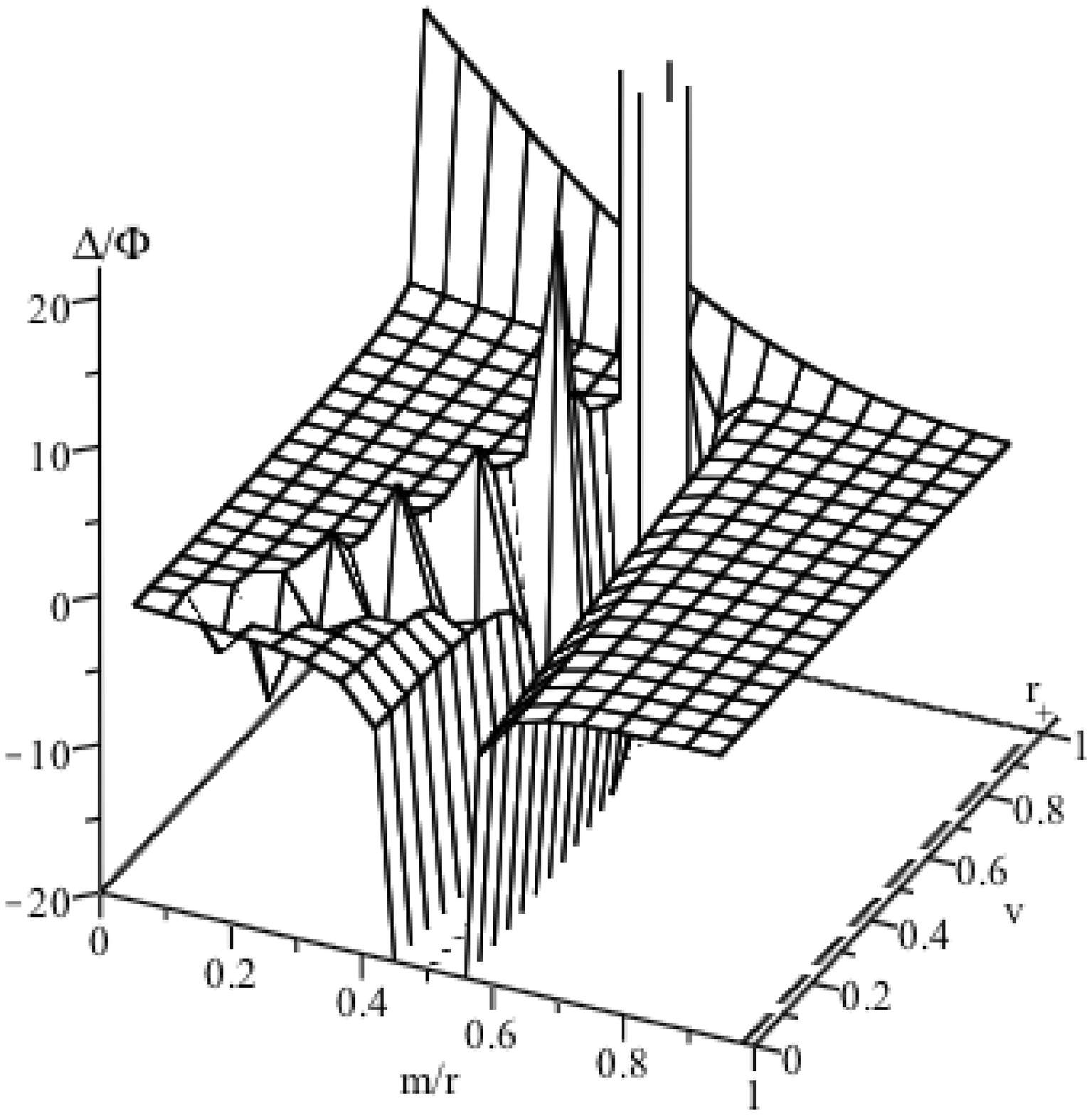}
\small{b) $a=0.99m$, $e=0.14m$ } }
\caption{
\small{The precession angle $\Delta / \Phi$ for a Kerr-Newman black hole for a pair of values of $a$ and $e$, that keep $r_+$ constant (dashed line). The dotted line represents the static limit surface on equatorial plane. The plots are asymptotic at $r=r_{st}$ and along a path $v=v_{fd}$. The peaks represent an asymptotic infinite wall.}}
\label{fig:Kerr-Newman_precession}
\end{figure}

For Region III the static limit (\ref{eq:static_limit}) is reduced to $r_{st}=m+\sqrt{m^2-e^2}$ on the equator. It coincides with the horizon for a Reissner-Nordstr\"{o}m
spacetime. The static limit $r_{st}$ represents again an asymptotic limit for
calculation of the precession angle $\Delta$ in the Kerr-Newman black
holes. Contrary to Kerr spacetime where the static limit is placed at $r=2m$, now is below and this limit depends in the charge of black hole, depicted by a dotted line in Fig.~\ref{fig:Kerr-Newman_precession}. In this figure, a) and b) have the same horizon (\ref{eq:horizons}), as well c) and d) between them. In figure a) the static limit is almost equal the horizon.

Once again, we can observe the infinite wall path due the coupling of $v_\mathrm{EPR}$ with $v_{fd}$. This asymptotic path is not now constrained  to the region $r>2m$ nor the horizon, but by the static limit.

Like in Kerr spacetime, the region between $r_+ \leqslant r < 2m$ is not affected by the frame-dragging. $\Delta$ tends asymptotically to infinity near $r_{st}$ and the coordinate system breakdown when $r$ equals $r_+$.

\subsection{Uncertainties in observers' positions and Bell's inequality}

When the hovering observers measure the spin of each entangled
particle, in principle they could adjust the direction of measure by
($\ref{eq:KN_delta}$) to get the perfect anti-correlation. As we
have seen, near the static limit and infinite wall path the precession
angle $\Delta$ will change fast, making impossible to keep a
position $\Phi$ without uncertainty for the hovering observers. As
\cite{T-U} mentioned, the error of the angle $\Theta$ to keep
the perfect EPR correlation is given by
\begin{equation}
  \delta \Theta = \delta \Phi \left| 1 + \frac{\Delta}{\Phi} \right|,
\label{eq:uncertainty_Theta}
\end{equation}
and near these asymptotic limits, $\delta \Theta$ can be
much larger than $\pi$. Thus the hovering observers cannot
determine the directions of measurement clearly enough to extract
the EPR correlation. To utilize the EPR correlation for quantum
communication, $\delta \Phi$ and $r$ must satisfy at least
\begin{equation}
  \delta \Theta < \pi \left| 1 + \frac{\Delta}{\Phi} \right|^{-1}.
\label{eq:uncertainty_Phi}
\end{equation}

When $r_{st}$ is reached for the Kerr-Newman spacetime, $\delta
\Phi$ must vanish because the velocity of the spin precession
(\ref{eq:general_velocity}) is infinite due the frame-dragging.
Therefore, on the static limit the hovering observers will not
obtain the right EPR correlation for the particles, unless they can
keep their positions $\Phi$ without uncertainty.

From the perspective of Bell's inequality, our previous results give
rise to a decrement in the degree of the violation of that
inequality, that is
\begin{equation}
\langle \mathcal{Q'S'} \rangle + \langle \mathcal{R'S'} \rangle + \langle \mathcal{R'T'} \rangle - \langle \mathcal{Q'T'} \rangle = 2\sqrt{2} \cos^2 \Delta,
\label{eq:Bell}
\end{equation}
where the trivial rotations of the local inertial frames $\pm \Phi$
has been discarded, the spin component of one particle is measured
in the $(\cos \Phi,~0,~-\sin \Phi)$ direction (component
$\mathcal{Q'}$) or in the (0, 1, 0) direction (component
$\mathcal{R'}$), and the spin component of the other is measured in
the $(\frac{-\cos \Phi}{\sqrt{2}},~\frac{-1}{\sqrt{2}},~\frac{-\sin \Phi}{\sqrt{2}})$ direction (component
$\mathcal{S'}$) or in the $(\frac{\cos \Phi}{\sqrt{2}},~\frac{-1}{\sqrt{2}},~\frac{-\sin \Phi}{\sqrt{2}})$
direction  (component $\mathcal{T'}$) as in Ref.~\cite{T-U} were established.

It is not possible to have local realistic theories as EPR demanded
because $\Delta$ in Eq.~(\ref{eq:Bell}) are computed from
local unitary operations. The angle $\Delta$ must vanish for maximal violation
of Bell's inequality, but as in previous sections we have seen, in
almost all Regions the spin precession has a non-vanishing value. Therefore
(\ref{eq:Bell}) will have a decrement in the degree of the violation
of the inequality, and the hovering observers must select different
set of directions of spin measure in order to get the maximal
violation of Bell's inequality. Then, these observers must take into
account the general relativistic effects arising from the dynamics
of the particles  and gravity. That is, spin component of one
particle must be measured in the $(\cos \Theta,~0,~-\sin \Theta)$
direction or in the (0, 1, 0) direction in the local inertial frame
at $\phi = \Phi$, and the spin component of the other one must be
measured in  the $(\frac{\cos \Theta}{\sqrt{2}},~\frac{-1}{\sqrt{2}},~\frac{-\sin \Theta}{\sqrt{2}})$
direction or in the $(\frac{\cos \Theta}{\sqrt{2}},~\frac{-1}{\sqrt{2}},~\frac{\sin \Theta}{\sqrt{2}})$
direction in the local inertial frame at $\phi = -\Phi$.

In a real situation, as soon as we reach the asymptotic limits, it will be almost impossible to keep the position $\Phi$ of
the hovering observers without a small uncertainty $\delta \Phi$,
which translates to an uncertainty in $\delta \Theta$ from
(\ref{eq:uncertainty_Theta}). This error in $\Theta$ decreases the
degree of violation as $2\sqrt{2}\cos^2 \delta \Theta$, and this error must be greater than 2 in order to restore the maximal
violation of Bell's inequality. Thus, from
(\ref{eq:uncertainty_Theta}), $\delta \Phi$ and $r$ must be adjusted
to
\begin{equation}
  \delta \Phi < \sqrt{2}\left| 1 + \frac{\Delta}{\Phi} \right|^{-1}.
\label{eq:uncertainty_Phi2}
\end{equation}

It is important to note that near the horizon for Reissner-Nordstr\"{o}m and the static limit and infinite wall path for Kerr and Kerr-Newman, the precession angle is asymptotically divergent and there is no
possible the maximal violation of Bell's inequality. Thus different
observers and vierbeins must be chosen to avoid this divergence of
spin precession angle.

\section{Conclusions}
\label{conclusions}

In this work we showed than when rotating black hole is considered, the frame-dragging must be introduced for the complete description of this spacetime. Then an additional velocity over particles must be relativistically incorporated.

The hovering observers were considered in order to have fixed reference frames that ensure reliable directions to compare the 1/2-spin quantum states. The total velocity measured by these observers was
performed as the addition of velocity of a ZAMO, plus the local velocity of the particles measured by the ZAMO.

These ZAMOs co-rotates the black hole due the frame-dragging and were used as a preliminary step before calculating the total local inertial velocity measured by the hovering observer.

We found that relativistic particles in black holes are difficult to keep in perfect anti-correlation  due to dynamical and gravitational effects,  as previous works showed \cite{a-m, t-u_sr1,T-U}.  The more parameters are added to the black holes, the more rich structure has the spacetime. Thus for the Kerr-Newman black hole, the spin precession angle (\ref{eq:KN_delta}) is obtained and  presents new features that differ considerably from the Schwarzschild \cite{T-U} and Kerr-Newman \cite{S-A} spacetimes.

Setting the parameter $a$ and $e$ to zero the spin precession angle for the Schwarzschild case is recovered \cite{T-U}. In the limit $r \to \infty$, those parameters vanish and the results for the Minkowski and Schwarzschild cases are also recovered. The new effects appear near the black hole horizon in the
Reissner-Nordstr\"{o}m case and the static limit $r_{st}$ for the Kerr-Newman spacetime.

Even that the total electric charge in real black holes should be zero, it was considered as an arbitrary parameter in order to illustrate its effect on the spin precession. The electromagnetic interaction between charged particles and charged black hole was not taken into account and remain to be explored in a future work.

Comparing with the Schwarzschild spacetime, the electric charge parameter shift the event horizon from $r=2m$ to $r=r_+$ in the Reissner-Nordstr\"{o}m case.

The angular momentum parameter establishes the static limit surface on the Kerr spacetime, where various interesting physical processes occur: it coincides with the Schwarzschild radius and represents one limit for calculation of the precession angle. The frame-dragging there has the maximal value, making massive particles ultra-relativistic and the spin precession angle $\Delta \to \infty$.

A remarkable difference was found when particles are close to the
rotating black hole event horizon. The precession angle is well defined, which contrasts with Scharzschild and Reissner-Nordstr\"{o}m cases.

Another new effect occurs when the velocity of the particles due the EPR process coincides with the frame-dragging velocity. One of the particles keeps their position relative to the hovering observer, meanwhile the other particle reach one observer. Then, the other observer does not measure anything and the spin precession angle tends to infinity.

It still possible to find circular orbits with perfect anti-correlation for $a$ and $e$ parameters, by fine-tuning of
velocity and position, in order to utilize the nonlocal correlation for quantum communication.  Moreover, when angular momentum parameter $a$ is considered, it is possible to keep a perfect anti-correlation close the static limit with three values of the local velocity due EPR process. This effect is not present in the Schwarzschild and Reissner-Nordstr\"{o}m cases.

In the Kerr-Newman spacetime, the static limit coincides with the horizon of the Reissner-Nordstr\"{o}m spacetime.

The paper showed that the choices of four-velocity, vierbein and observers are important for the ability to communicate non-locally in a Kerr-Newman spacetime using an EPR pair of spinning particles. If the spins are measured in appropriately chosen different directions, we can obtain the perfect anti-correlation and the maximal violation of Bell's inequality. But as soon as the particles get closer to the asymptotic limits, their velocities increase very fast until asymptotically reach speed of light, with a rapid spin precession. The hovering observers would not be able to adjust the direction of measure of the spin, making virtually  impossible the quantum communication.

It would be interesting to make an analytic continuation in order to find the interior solution supporting the spin precession below the event horizon $r_+$. Also one could consider a gedanken experiment mounted on ZAMOs for the Kerr-like spacetime.

Future work will extend the analysis discussed in the present paper to evaluate the behavior of spin precession on any general type D metric, making possible to investigate the EPR anti-correlation of
particles in a a seven parametric Pleba\'{n}ski-Demia\'{n}ski spacetime  \cite{Plebanski:1976gy}.

\vskip 1truecm
\centerline{\bf Acknowledgments} It is a pleasure to thank the
referee of QIC journal for very useful suggestions. We also thank R.
Morones, J. Comparan and the staff of the FCFM, UANL for kind
support. The work of F. R.-P. was supported by a CONACyT graduate
fellowship. The work of  H. G.-C. was supported in part by a CONACyT
grant 128761.


\end{document}